\documentclass[preprint,amsmath,amssymb,prd]{revtex4}

\def\al{\alpha}
\def\be{\beta}
\def\ga{\gamma}
\def\de{\delta}
\def\ep{\epsilon}

\def\et{\eta}

\def\ka{\kappa}
\def\la{\lambda}

\def\rh{\rho}

\def\si{\sigma}

\def\ta{\tau}

\def\ph{\phi}

\def\ps{\psi}

\def\Ga{\Gamma}

\def\La{\Lambda}

\def\cE{{\cal E}}
\def\cl{{\cal L}}
\def\cL{{\cal L}}
\def\cO{{\cal O}}

\def\fr#1#2{{{#1} \over {#2}}}
\def\half{{\textstyle{1\over 2}}}

\def\frac#1#2{{\textstyle{{#1}\over {#2}}}}

\def\vev#1{\langle {#1}\rangle}

\def\lsim{\mathrel{\rlap{\lower4pt\hbox{\hskip1pt$\sim$}}
    \raise1pt\hbox{$<$}}}
\def\gsim{\mathrel{\rlap{\lower4pt\hbox{\hskip1pt$\sim$}}
    \raise1pt\hbox{$>$}}}
\def\sqr#1#2{{\vcenter{\vbox{\hrule height.#2pt
         \hbox{\vrule width.#2pt height#1pt \kern#1pt
         \vrule width.#2pt}
         \hrule height.#2pt}}}}

\def\tr#1{{\rm{tr}}\left[ #1 \right]}
\def\ttr#1{{\rm{tr}}[#1]}

\def\det#1{{\rm{det}}\left[ #1 \right]}

\def\prt{\partial}

\def\pt#1{\phantom{#1}}

\def\ol#1{\overline{#1}}

\def\gam#1#2#3{\Ga^{#1}_{{\pt{#1}}{#2}{#3}}}
\def\gamu#1#2#3{\Ga^{#1\pt{#2}#3}_{{\pt{#1}}{#2}}}
\def\gaml#1#2#3{\Ga^{\pt{{#1}{#2}}#3}_{{#1}{#2}}}

\def\ab{{\al\be}}
\def\cd{{\ga\de}}
\def\mn{{\mu\nu}}

\def\fg{\frak{g}}
\def\fh{\frak{h}}

\def\fc{\frak{c}}
\def\fC{\frak{C}}
\def\fK{\frak{K}}
\def\fM{\frak{M}}

\def\fN{\frak{N}}
\def\fV{\frak{V}}
\def\fx{\frak{x}}
\def\fX{\frak{X}}
\def\fy{\frak{y}}
\def\fY{\frak{Y}}

\def\tC{{\widetilde C}}
\def\tfC{{\widetilde \fC}}
\def\tfX{{\widetilde \fX}}

\def\tg{{N}}
\def\ftg{{\fN}}
\def\tO{{O}}
\def\tm{{M}}
\def\ftm{{\fM}}
\def\mm{{m}}

\def\tc{{c}}
\def\kk{{K}}
\def\cc{{c}}

\newcommand{\beq}{\begin{equation}}
\newcommand{\eeq}{\end{equation}}
\newcommand{\bea}{\begin{eqnarray}}
\newcommand{\eea}{\end{eqnarray}}
\newcommand{\bit}{\begin{itemize}}
\newcommand{\eit}{\end{itemize}}
\newcommand{\rf}[1]{(\ref{#1})}

\begin{document}

\title{Gravity from spontaneous Lorentz violation}

\author{V.\ Alan Kosteleck\'y$^a$ and Robertus Potting$^b$}

\affiliation{
$^a$Physics Department, Indiana University,
Bloomington, IN 47405
\\
$^b$CENTRA, Physics Department, FCT, Universidade do Algarve,
         8000-139 Faro, Portugal
}

\date{IUHET 523, January 2009}

\begin{abstract}
We investigate a class of theories 
involving a symmetric two-tensor field
in Minkowski spacetime with a potential triggering
spontaneous violation of Lorentz symmetry.
The resulting massless Nambu-Goldstone modes
are shown to obey the linearized Einstein equations in a fixed gauge.
Imposing self-consistent coupling to the energy-momentum tensor
constrains the potential for the Lorentz violation.
The nonlinear theory generated  
from the self-consistent bootstrap
is an alternative theory of gravity,
containing kinetic and potential terms 
along with a matter coupling.
At energies small compared to the Planck scale, 
the theory contains general relativity,
with the Riemann-spacetime metric constructed
as a combination of the two-tensor field and the Minkowski metric.
At high energies,
the structure of the theory is qualitatively different 
from general relativity.
Observable effects can arise 
in suitable gravitational experiments.

\end{abstract}

\maketitle

\section{Introduction}

The idea that physical Lorentz symmetry
could be broken in a fundamental theory of nature
has received much attention in recent years.
One attractive mechanism is spontaneous Lorentz violation,
in which an interaction drives an instability 
that triggers the development of nonzero vacuum values 
for one or more tensor fields
\cite{ks}.
Unlike explicit breaking,
spontaneous Lorentz violation is compatible with
conventional gravitational geometries
\cite{akgrav},
and it is therefore advantageous for model building.
However,
spontaneous violation of a continuous global symmetry
comes with massless excitations,
the Nambu-Goldstone (NG) modes
\cite{ng}.
Among the challenges facing attempts to construct
realistic models with spontaneous Lorentz violation 
is accounting for the role of the corresponding NG modes
and interpreting them phenomenologically.

Since the NG modes are intrinsically massless,
they can generate long-range forces.
One intriguing possibility is that 
they could reproduce one of the long-range forces
known to exist in nature. 
For electrodynamics,
for example,
the Einstein-Maxwell equations in a fixed gauge
naturally emerge from the NG sector
of certain gravitationally coupled vector theories 
with spontaneous Lorentz violation
known as bumblebee models
\cite{ks2,bk}.
For gravity itself,
the gravitons can be interpreted as the NG modes
from spontaneous Lorentz violation
in several ways.
As fundamental field excitations,
gravitons can be identified with the NG modes 
of a symmetric two-tensor field $C^\mn$
in a theory with a potential inducing spontaneous Lorentz violation,
which generates the linearized Einstein equations in a fixed gauge
\cite{kp}.
Alternatively,
gravitons as composite objects can be understood 
as the NG modes of spontaneous Lorentz violation
arising from self interactions of 
vectors
\cite{vector},
fermions
\cite{fermion}, 
or scalars
\cite{scalar},
following related ideas for composite photons
\cite{photons}.
Other interpretations of the NG modes include
a new spin-dependent interaction
\cite{ahclt}
and various new spin-independent forces
\cite{kt}.
For certain theories in Riemann-Cartan spacetimes, 
the NG modes can instead be absorbed 
into the spin connection via the Lorentz-Higgs effect
\cite{bk}. 

In the present work,
we investigate the possibility that
the full nonlinear structure of general relativity 
can be recovered from an alternative theory of gravity 
with spontaneous Lorentz violation
in which the gravitons are fundamental excitations
identified with the NG modes. 
General relativity has the interesting feature
that it can be reconstructed uniquely from massless spin-2 fields
by requiring consistent self-coupling
to the energy-momentum tensor
\cite{bootstrap,sd,sd2,sd3,sd4}.
For example,
the linearized theory describing gravitational waves
via a symmetric two-tensor $h^\mn$
propagating in a spacetime with Minkowski metric $\et_\mn$
contains sufficient information
to reconstruct the full nonlinearity of general relativity
when self consistency is imposed.
Here, 
we demonstrate that applying this bootstrap method 
to a linearized theory with a symmetric two-tensor field $C^\mn$
and a potential $V(C^\mn, \et_\mn )$
inducing spontaneous Lorentz violation
yields an alternative theory of gravity,
which we call cardinal gravity
\cite{foot1}.
The coupling of the cardinal field to the matter sector
is derived,
and constraints from existing experiments are considered. 
We show that the action of cardinal gravity 
corresponds to the Einstein-Hilbert action at energies
small compared to the Planck scale.
However,
the structure of the theory at high energies 
is qualitatively different from that of general relativity.
Our results indicate that cardinal gravity 
is a viable alternative theory of gravity
exhibiting some intriguing features
in extreme gravitational environments.

We begin this work in Sec.\ \ref{Linearized analysis}
by presenting the linearized cardinal theory 
and a discussion of 
its correspondence to linearized general relativity.
Section \ref{Bootstrap procedure}
reviews the bootstrap procedure for general relativity
and obtains some generic results.
For general relativity,
the bootstrap procedure yields a unique answer
even if a potential for $h^\mn$ is allowed
\cite{sd3}.
In the context of spontaneous Lorentz violation,
the phase transition circumvents this uniqueness.
However,
the nontrivial integrability conditions
required for implementing the bootstrap
constrain the form of the potential $V$.
In Sec.\ \ref{Bootstrap for cardinal gravity},
we obtain differential equations expressing
the integrability conditions
and derive acceptable potentials $V$.
This section also applies the bootstrap
to yield the full cardinal gravity.
Certain aspects of the extrema of the potential are considered, 
and alternative bootstrap procedures
are discussed.
The coupling of the cardinal field to the matter sector
and some experimental implications
are studied in Sec.\ \ref{Coupling to matter}.
A summary of the results 
and a discussion of their broader implications
is provided in Sec.\ \ref{Discussion}.
Throughout this work,
we use the conventions of Ref.\ \cite{akgrav}.

\section{Linearized analysis}
\label{Linearized analysis}

In this section,
the linear cardinal theory is defined and investigated.
We show that its NG sector
is equivalent to conventional linearized gravity 
in a special gauge. 

\subsection{Linear cardinal theory}

Consider first the action 
for the symmetric two-tensor cardinal field $C^\mn$
defined in a background spacetime.
For definiteness and simplicity,
we take the background 
to be Minkowski spacetime with metric $\et_\mn$,
although a more general background
could be countenanced and treated with similar methods.
We suppose the kinetic term in the action 
is quadratic in $C^\mn$,
so the derivative operators in the equation of motion
are linear in $C^\mn$.
The NG excitations of $C^\mn$ 
subsequently play the role of the metric fluctuation 
in a linearized theory of gravity.
The action is assumed to generate 
spontaneous violation of Lorentz symmetry
through a potential $V(C^{\mu\nu},\et_{\mu\nu})$.  

\subsubsection{Basics}

The Lagrange density for the linear cardinal theory 
is taken to be 
\bea
\cl_C &=&
\half
C^{\mu\nu} K_{\mu\nu\al\be} C^{\al\be} - V(C^{\mu\nu},\et_{\mu\nu}).
\label{clag}
\eea
Here,
$K_{\mu\nu\al\be}$ is the usual quadratic kinetic operator 
for a massless spin-2 field.
Allowing for an arbitrary scaling parameter $\ka$
to be chosen later,
$K_{\mu\nu\al\be}$ can be written in cartesian coordinates as
\bea
K_{\mu\nu\al\be} &=&
\half \ka[
(-\et_{\mu\nu}\et_{\al\be} 
+ \half \et_{\mu\al}\et_{\nu\be}
+ \half \et_{\mu\be}\et_{\nu\al}) 
\prt^\la \prt_\la
\nonumber\\
&&
+\et_{\mu\nu} \prt_\al\prt_\be
+\et_{\al\be} \prt_\mu\prt_\nu
\nonumber\\
&&
-\half \et_{\mu\al} \prt_\nu\prt_\be
-\half \et_{\nu\al} \prt_\mu\prt_\be 
\nonumber\\
&&
-\half \et_{\mu\be} \prt_\nu\prt_\al
-\half \et_{\nu\be} \prt_\mu\prt_\al ],
\label{clagK}
\eea
where $\et_{\mu\nu}$ is the Minkowski metric 
with diagonal entries $(-1, 1, 1, 1)$
as the only nonzero components.
As usual,
in other coordinate systems
the Minkowski metric takes different forms 
and covariant derivatives must be used.
The equations of motion obtained by varying Eq.\ \rf{clag}
with respect to $C^\mn$ are
\bea
K_{\mu\nu\al\be} C^{\al\be} - \fr{\de V}{\de C^{\mu\nu}} &=& 0.
\label{ceqmot}
\eea

The theory \rf{clag} has various symmetries.
It is invariant under translations
and under global Lorentz transformations.
For infinitesimal parameters $\ep_{\mu\nu}=-\ep_{\nu\mu}$,
the latter take the form
\bea
C^{\mu\nu}
&\rightarrow&
C^{\mu\nu}
+ \ep^\mu_{\pt{\mu}\al} C^{\al\nu}
+ \ep^\nu_{\pt{\mu}\al} C^{\al\mu},
\nonumber\\
\et_{\mu\nu}
&\rightarrow&
\et_{\mu\nu} .
\label{LTs}
\eea
There are also local spacetime symmetries,
including invariance under local Lorentz transformations
on the tangent space at each point 
and invariance under diffeomorphisms of the Minkowski spacetime.
These local symmetries play a subsidiary role 
in the present context.

In addition to the spacetime symmetries,
the form of the kinetic operator \rf{clagK}
ensures that the kinetic term is by itself invariant 
under gauge transformations of $C^{\mu\nu}$ alone,
\bea
C^{\mu\nu}
&\rightarrow&
C^{\mu\nu} - \prt^\mu \La^\nu - \prt^\nu \La^\mu,
\nonumber\\
\et_{\mu\nu}
&\rightarrow&
\et_{\mu\nu} .
\label{gauge}
\eea
However,
one or more of these four gauge symmetries may be
explicitly broken by the potential $V$,
so the Lagrange density \rf{clag} 
contains between zero and four gauge degrees of freedom
depending on the choice of $V$.
Since $C^{\mu\nu}$ has ten independent components,
it follows that there are 
between six and ten physical or auxiliary fields.

The potential $V$ for the theory \rf{clag}
is a scalar function of
the cardinal field $C^{\mu\nu}$
and the Minkowski metric $\et_{\mu\nu}$.
The only scalars 
that can be formed from these two objects
involve traces of products of the combination 
$C^{\mu\al} \et_{\al\nu}$.
The scalar $X_m$ with $m$ such products has the form
\beq
X_m = \tr{(C \et)^m}.
\eeq
Here, 
we have introduced a convenient matrix notation 
$(C \et)^\mu_{\pt{\mu}\nu} \equiv 
C^{\mu\al}\et_{\al\nu}$.
Since $C \et$ is a symmetric 4$\times$4 matrix,
there are at most four independent scalars $X_m$,
so we can restrict attention to the cases $X_m = 1,2,3,4$.
It follows that the potential $V$ can be written as
\beq
V = V(X_1, X_2, X_3, X_4)
\eeq
without loss of generality.
For definiteness,
$V$ is assumed to be positive everywhere
except at its absolute minimum,
which is taken to be zero. 

Under the gauge transformation \rf{gauge},
each scalar $X_m$ transforms nontrivially 
and therefore explicitly breaks one symmetry. 
For simplicity in what follows,
we assume the potential $V$ depends on all four 
independent scalars $X_m$, 
so the gauge symmetry \rf{gauge} is completely broken
for generic field configurations.
With this assumption,
the theory describes ten physical or auxiliary fields
and zero gauge fields.
This assumption could be relaxed,
but the resulting discussion would involve
additional gauge-fixing considerations. 

The potential $V$
is taken to have a minimum in which 
$C^{\mu\nu}$ attains
a nonzero vacuum value 
\beq
\vev{C^{\mu\nu}} \equiv c^{\mu\nu}.
\label{vev2}
\eeq
In this minimum,
the scalars $X_m$ have vacuum values
\beq
\vev{X_m} \equiv x_m = \tr{(c \et)^m}.
\label{vacsol}
\eeq
These vacuum values spontaneously break particle Lorentz symmetry,
but they leave unaffected the structure of 
observer Lorentz and general coordinate transformations,
which amount to coordinate choices.
To avoid complications with soliton-type solutions,
we also suppose $c^{\mu\nu}$ is constant,
\beq
\prt_\al c^{\mu\nu} = 0
\label{constc}
\eeq
in cartesian coordinates.

Given a vacuum value $c^{\mu\nu}$,
the freedom of coordinate choice can be used to
adopt a canonical form.
For definiteness and simplicity,
we assume in what follows that
the matrix 
$(c \et)^\mu_{\pt{\mu}\nu} \equiv c^{\mu\al} \et_{\al\nu}$
has four inequivalent nonzero real eigenvalues.
This implies,
for example,
invertibility and the existence
of one timelike and three spacelike eigenvectors.
It also implies that all six Lorentz transformations
are spontaneously broken.
The consequences of other possible choices
may differ from the discussion below 
and would be interesting to explore,
but they lie beyond our present scope.

\subsubsection{Nambu-Goldstone and massive modes}
\label{Nambu-Goldstone and massive modes}

The physical degrees of freedom 
contained in the cardinal field $C^{\mu\nu}$ 
can be taken as fluctuations 
about the vacuum value $c^{\mu\nu}$.
We write
\beq
C^{\mu\nu} = c^{\mu\nu} + \tC^{\mu\nu}.
\label{flucts}
\eeq
The fluctuation field $\tC^{\mu\nu}$ is symmetric 
and has ten independent components,
which include both the NG modes and the massive modes
in the theory.

To identify the NG modes,
we can make 
virtual infinitesimal symmetry transformations 
using the broken generators
acting on field vacuum values,
and then promote the corresponding parameters
to field excitations.
An infinitesimal Lorentz transformation
with parameters $\ep_{\mu\nu}=-\ep_{\nu\mu}$ yields 
\beq
\vev{C^{\mu\nu}} \to 
c^{\mu\nu} 
+ \ep^\mu_{\pt{\mu}\al} c^{\al\nu}
+ \ep^\nu_{\pt{\mu}\al} c^{\al\mu}.
\label{TLTchange}
\eeq
Since there are six Lorentz transformations
(three rotations and three boosts),
there could in principle be up to six Lorentz NG modes,
corresponding to the promotion of the six parameters 
$\ep_{\mu\nu}$
to fields
$\cE_{\mu\nu}= -\cE_{\nu\mu}$
\cite{bk,bfk}.
For $c^{\mu\nu}$ satisfying our assumed conditions,
the maximal set of six NG modes appears.
In general,
the NG modes in $\tC^{\mu\nu}$ are contained 
in the fluctuations $\tg^{\mu\nu}$ defined by
\bea
\tC^{\mu\nu} \supset \tg^{\mu\nu}
&=& \cE_{\pt{al}\al}^{\mu} c^{\al\nu}
+ \cE_{\pt{al}\al}^{\nu} c^{\al\mu}
\nonumber\\ 
&\equiv& \tO^{\mu\nu\al\be} \cE_{\al\be},
\label{eps}
\eea
where
\beq
\tO^{\mu\nu\al\be} = 
\half 
( \et^{\mu\al} c^{\nu\be} 
+ \et^{\nu\al} c^{\mu\be} 
- \et^{\mu\be} c^{\nu\al}
- \et^{\nu\be} c^{\mu\al} 
).
\label{M}
\eeq
Since there are six independent fields in $\cE_{\mu\nu}$,
the ten symmetric components of $\tg^{\mu\nu}$
must obey four identities.
For $c^{\mu\nu}$ satisfying our assumed conditions,
we find these identities can be expressed as 
\beq
\tr{\tg \et (c \et)^j} = 0,
\label{decid}
\eeq
with $j = 0,1,2,3$.

In addition to the six NG modes in the field $\tg^{\mu\nu}$,
the fluctuation $\tC^{\mu\nu}$ includes four massive modes.
These are contained in the field $\tm^{\mu\nu}$
given by
\beq
\tm^{\mu\nu} = \tC^{\mu\nu} - \tg^{\mu\nu},
\eeq
subject to a suitable orthogonality condition.
The symmetric field $\tm^{\mu\nu}$ has ten components
but only four independent degrees of freedom,
which we denote here by $\mm_j$, $j = 0,1,2,3$.
For some purposes,
it is convenient to expand $\tm^{\mu\nu}$ as
\beq
\tm^{\mu\nu} =
\mm_0 \et^{\mu\nu}
+ \mm_1 c^{\mu\nu}
+ \mm_2 (c \et c)^{\mu\nu}
+ \mm_3 (c \et c \et c)^{\mu\nu}.
\label{msinM}
\eeq
The fields $\tg^{\mu\nu}$ and $\tm^{\mu\nu}$
obey identities expressing a kind of orthogonality:
\beq
\tr{\tg \et (\tm \et)^j} = 0,
\label{demid}
\eeq
with $j = 0,1,2,3$.
More generally,
we find
\beq
\tr{\tg \et ~F(\tc\et, \tm\et)} = 0,
\label{arbdemid}
\eeq
where $F(\tc\et, \tm\et)$ is 
an arbitrary matrix polynomial in $\tc\et$ and $\tm\et$.

With the expansion \rf{msinM},
the fluctuation $\tC^{\mu\nu}$ can be written 
\bea
\tC^{\mu\nu} 
&=& \tg^{\mu\nu} 
+ \sum_{j=0}^{3} \mm_j [(c \et)^j]^\mu_{\pt{\mu}\al} \et^{\al\nu} .
\eea 
Using this equation,
the four massive modes $\mm_j$ 
can be expressed in terms of $\tC^{\mu\nu}$.
Multiplying by 
$[\et (c\et)^k]_{\mu\nu}$ 
with $k = 0,1,2,3$
and applying the identities \rf{decid}
yields the four equations
\bea
&
\left(
\begin{array}{cccc}
4 & \ttr{c\et} & \ttr{(c\et)^2} &  \ttr{(c\et)^3}
\\
\ttr{c\et} & \ttr{(c\et)^2} &  \ttr{(c\et)^3} & \ttr{(c\et)^4}
\\
\ttr{(c\et)^2} & \ttr{(c\et)^3} & \ttr{(c\et)^4} & \ttr{(c\et)^5}
\\
\ttr{(c\et)^3} & \ttr{(c\et)^4} & \ttr{(c\et)^5} & \ttr{(c\et)^6}
\end{array}
\right)
\left(
\begin{array}{c}
\mm_0 
\\
\mm_1 
\\
\mm_2 
\\
\mm_3 
\\
\end{array}
\right)
\nonumber\\
&
\hskip 140 pt
= \left(
\begin{array}{c}
\ttr{\tC \et} 
\\
\ttr{\tC \et (c \et)} 
\\
\ttr{\tC \et (c \et)^2} 
\\
\ttr{\tC \et (c \et)^3} 
\\
\end{array}
\right).
\nonumber\\
\label{meq}
\eea
The traces $\ttr{(c\et)^p}$ with $p = 5,6$
can be rewritten in terms of $\ttr{(c\et)^m}$ 
with $m = 1,2,3,4$
using the Hamilton-Cayley theorem.
In terms of the eigenvalues $\cc_j$ of the matrix $c\et$,
the determinant of the 4$\times 4$ matrix $\cO$
on the left-hand side takes the form
\bea
\det\cO &=& 
\prod_{{\scriptstyle j,k = 0}\atop{\scriptstyle j<k}}^3
(\cc_j - \cc_k)^2 .
\eea
For the matrix $c\et$ satisfying our assumed conditions,
it follows that Eq.\ \rf{meq} can be inverted 
to give explicit expressions
for each of the four massive modes $\mm_j$
in terms of $\tC^{\mu\nu}$.
These somewhat lengthy expressions involve
the four field traces $\ttr{\tC \et (c \et)^j}$
with $j = 0,1,2,3$
and the four quantities $\ttr{(c\et)^m}$
with $m = 1,2,3,4$.
Their explicit forms are unnecessary 
in the discussion that follows,
so we omit them here.

The above considerations reveal that the decomposition 
of the cardinal field $C^{\mu\nu}$
in terms of NG and massive modes is
\beq
C^{\mu\nu} = c^{\mu\nu} + \tg^{\mu\nu} + \tm^{\mu\nu} .
\label{decomp}
\eeq
The potential $V$ can therefore be viewed as a function
of $\tg^{\mu\nu}$ and $\tm^{\mu\nu}$ with constraints
added to restrict these fields
to their independent degrees of freedom,
or equivalently
as a function of the Lorentz NG modes $\cE_{\mu\nu}$
and the massive modes $\mm_j$:
\bea
V(C^{\mu\nu},\et_{\mu\nu}) &=& 
V(c^{\mu\nu},\cE_{\mu\nu},\mm_0,\mm_1,\mm_2,\mm_3,\et_{\mu\nu}) .
\quad
\label{vcepm}
\eea

To investigate the correspondence of the linear
cardinal theory \rf{clag} to linearized general relativity,
it is useful to restrict attention to the pure NG sector.
This can be achieved by considering the limit 
of infinite mass for the fields $\mm_j$.
Alternatively,
the potential $V$ can be replaced with 
the Lagrange-multiplier limit $V_\la$
given by
\beq
V_\la =  \sum_{m=1}^4 \la_m (X_m - x_m) ,
\label{vlm}
\eeq
where the quantities $\la_m$ 
are four Lagrange-multiplier fields.
This potential freezes all fluctuations of $C^{\mu\nu}$
away from the potential minimum.
In this limit,
the independent degrees of freedom
in the field fluctuations $\tC^{\mu\nu}$
are therefore restricted to the NG modes
$\cE_\mn$ or, 
equivalently,
$\tC^{\mu\nu} \to \tg^{\mu\nu}$
subject to the constraints \rf{decid}.
If desired,
the on-shell values of $\la_j$
can be set to zero 
by a suitable choice of initial conditions.
Equivalent results could be obtained 
via an alternative Lagrange density
involving a potential $V$ with 
quadratic Lagrange-multiplier terms instead
\cite{bfk}.
In any event,
if the graviton is to be identified
with the Lorentz NG modes in the theory \rf{clag},
it follows that the field $\tg^{\mu\nu}$ must be 
the candidate graviton field.

\subsubsection{Equations of motion for NG modes}

The behavior of the candidate graviton field $\tg^{\mu\nu}$
is determined by its equations of motion.
In the pure NG sector with vanishing Lagrange multipliers,
the theory \rf{clag} with the potential \rf{vlm}
is equivalent to an effective Lagrange density 
$\cl_{\rm NG}$
for the independent degrees of freedom,
which are the Lorentz NG modes $\cE_{\mu\nu}$.
We can therefore write
\bea
\cl_{\rm NG} &=&
\half 
\tO^{\mu\nu\rh\si} \cE_{\rh\si} 
K_{\mu\nu\al\be}
\tO^{\al\be\ga\de} \cE_{\ga\de} .
\label{eplag}
\eea
Varying $\cl_{\rm NG}$ with respect to
the independent degrees of freedom $\cE_{\mu\nu}$
yields the six equations of motion 
\beq
\tO^{\mu\nu\rh\si} 
K_{\mu\nu\al\be}
\tO^{\al\be\ga\de} \cE_{\ga\de} =0.
\label{epeqmot}
\eeq
These can equivalently be written as
\beq
\tO^{\mu\nu\rh\si} 
K_{\mu\nu\al\be}
\tg^{\al\be} =0,
\label{deceqmot}
\eeq
where the constraints \rf{decid} are understood.

To solve these equations we can use Fourier decomposition,
transforming to momentum space with 4-momentum $k_\mu$.
It is convenient to introduce the scalars $\kk_{m,n}$ and $\kk_m$,
defined by the matrix equations
\beq
\kk_{m,n} \equiv k(c \et)^m \tg \et (c \et)^n k,
\quad
\kk_m \equiv k(c \et)^mk.
\label{deckdef}
\eeq
Note that $\kk_{m,n} = \kk_{n,m}$
by virtue of the symmetry of $\tg^{\mu\nu}$.
Contraction of the equations of motion \rf{deceqmot}
with $k(c\et)^m$ yields the following results,
equivalent in content to the original equations of motion: 
\bea
&
\hskip -40pt
k^2 \kk_{m+1,n}
+ \kk_m \kk_{0,n+1}
+ \kk_{n+1} \kk_{m,0}
&
\nonumber\\
&-k^2 \kk_{m,n+1}
- \kk_n \kk_{m+1,0}
- \kk_{m+1} \kk_{0,n} 
= 0.
&
\label{ckeqmot} 
\eea
These expressions are solved by 
the on-shell condition $k^2 = 0$
and the constraint 
$k_\mu \tg^{\mu\nu} = 0$.
We have verified that no physical off-shell solutions exist.
The on-shell solutions are modes obeying
the usual massless wave equation,
\beq
\prt_\la\prt^\la \tg^{\mu\nu} = 0,
\label{redepeqmot}
\eeq
subject to the harmonic condition
\beq
\prt_\mu \tg^{\mu\nu}=0.
\label{harmonic}
\eeq
The latter imposes four constraints
on the six independent degrees of freedom in $\tg^{\mu\nu}$.

We thus see that only two combinations of 
the six massless Lorentz NG modes $\cE_{\mu\nu}$ propagate
as physical on-shell fields.
The other four NG modes are auxiliary.
With the full potential $V$
replaced by the Lagrange-multiplier limit $V_\la$,
the four Lagrange multipliers can be viewed as
playing a role analogous to that
of the four frozen massive modes $\mm_j$.

\subsection{Correspondence to linearized general relativity}

In this subsection,
we show the correspondence between 
the restriction of the linear cardinal theory to the NG sector
and the usual weak-field limit of general relativity 
describing a massless spin-2 graviton field
$h_{\mu\nu}$ propagating in a background Minkowski spacetime.

Consider the Lagrange density
for a free symmetric massless spin-2 field $h_{\mu\nu}$,
which is of the form \rf{clag}
with $C^{\mu\nu}$ replaced by $h^{\mu\nu}$
and the potential $V$ set to zero:
\beq
\cl_h =
\half 
h^{\mu\nu} K_{\mu\nu\al\be} h^{\al\be} .
\label{hlag}
\eeq
The definition of $K_{\mu\nu\al\be}$ in Eq.\ \rf{clagK}
implies
\beq
K_{\mu\nu\al\be} h^{\al\be} \equiv -\ka G^L_\mn,
\eeq
where $G^L_\mn$ is the Einstein tensor
linearized in $h^\mn$.
At this stage,
the value of $\ka$ can be fixed 
by requiring a match to the conventional normalization
of the linearized action for general relativity 
in the presence of a matter coupling given by
\beq
\cl^L_T = \half h^\mn T_{\rm M \mn},
\eeq
where $T_{\rm M \mn}$ is the matter energy-momentum tensor.
This match fixes $\ka$ to be 
\beq
\ka = \fr 1 {16\pi G_N},
\label{kappa}
\eeq
where $G_N$ is the Newton gravitational constant.

{\it A priori,}
$h^{\mu\nu}$ has ten degrees of freedom.
However,
the theory is invariant under the four gauge transformations
\bea
h^{\mu\nu}
&\rightarrow&
h^{\mu\nu} - \prt^\mu \xi^\nu - \prt^\nu \xi^\mu,
\label{hgauge}
\eea
so four gauge-fixing conditions can be imposed on $h^{\mu\nu}$.
Numerous choices of gauge appear in the literature.
For free wave propagation,
a common choice is transverse-traceless gauge,
which imposes 
\beq
n_\mu h{}^{\mu\nu}= 0,
\quad  
h \equiv h{}^\mu_{\pt{\mu}\mu} = 0, 
\eeq
for a unit timelike vector $n_\mu$.
For suitable initial conditions,
the harmonic condition
\beq
\prt_\mu h{}^{\mu\nu}=0
\label{harmonicx}
\eeq
then follows from the equations of motion.
However,
this gauge is not the only possible choice.
Here,
we demonstrate the existence of an alternative gauge condition 
on $h^{\mu\nu}$ that yields directly a match 
to the NG effective Lagrange density \rf{eplag}.

The conditions fixing this alternative `cardinal' gauge 
at linear order in $h^{\mu\nu}$ are 
\beq
\tr{h \et (c \et)^j} = 0, 
\label{newg}
\eeq
where $j = 0,1,2,3$.
In this expression,
$(c \et)^\mu_{\pt{\mu}\nu} \equiv c^{\mu\al} \et_{\al\nu}$
is a constant matrix assumed 
to have four inequivalent nonzero real eigenvalues,
which we denote by 
$\cc_j$, $j = 0,1,2,3$.
This assumption ensures the four conditions \rf{newg}
are independent.
For the present purpose of matching 
to the linear cardinal theory \rf{clag},
the quantity $c^{\mu\nu}$ is to be identified 
with the vacuum value of $C^{\mu\nu}$ in Eq.\ \rf{vev2},
so we denote it by the same symbol.

To show that the conditions \rf{newg} are indeed a choice of gauge,
we can consider an arbitrary initial field
$h^{\prime\mu\nu}$ 
and seek quantities $\xi^\mu$ 
such that a gauge transformation of the form \rf{hgauge}
generates the desired field $h^{\mu\nu}$ satisfying \rf{newg}.
In momentum space,
the gauge transformation \rf{hgauge}
takes the form 
\beq
h^{\mu\nu}= h^{\prime\mu\nu} -ik^\mu\xi^\nu - ik^\nu\xi^\mu.
\label{cond}
\eeq
The requirements on $\xi^\mu$ become
\bea
ik_\mu \xi^\mu 
&=&
\half \tr {\et h^\prime},
\nonumber\\
ik_\al (c\et)^\al_{\pt{\al}\mu} \xi^\mu 
&=&
\half \tr {\et h^\prime \et c},
\nonumber\\
ik_\al [(c\et)^2]{}^\al_{\pt{\be}\mu} \xi^\mu 
&=&
\half \tr {\et h^\prime (\et c)^2},
\nonumber\\
ik_\al [(c\et)^3]{}^\al_{\pt{\be}\mu} \xi^\mu 
&=&
\half \tr {\et h^\prime (\et c)^3}.
\label{gaugereq}
\eea
This represents a set of four equations 
for the four unknowns $\xi^\mu$,
which can be regarded as a matrix equation.
The set has a unique solution if the 
4$\times$4 matrix 
generated by the coefficients of $\xi^\mu$ is invertible.
Then,
the four 4-vectors 
$k_\mu$, 
$k_\al (c\et)^\al_{\pt{\be}\mu}$,
$k_\al [(c\et)^2]{}^\al_{\pt{\be}\mu}$,
$k_\al [(c\et)^3]{}^\al_{\pt{\be}\mu}$
are linearly independent,
and so 
\beq
\ep_{\mu\nu\rh\si}
k^\mu
k_\al c^{\al\nu}
k_\be (c\et c)^{\be\rh}
k_\ga (c\et c\et c)^{\ga\si}
\neq 0.
\label{detcond}
\eeq
Expanding the 4-vector $k$ in terms of the 
eigenvectors $e^{(a)}$ of the matrix $c\et$
shows that this condition is indeed satisfied
for generic $k$,
for which all components $k^{(a)} = k\cdot e^{(a)}$ are nonzero.
It follows that the cardinal gauge \rf{newg} 
can be attained everywhere 
in conventional linearized general relativity,
except for special $k$ at which additional gauge fixing is required.
This remnant gauge freedom is analogous 
to that of axial gauge in electrodynamics
\cite{axial}.
Similarly,
in the context of spontaneous Lorentz violation,
the linearized potential for the vector field
in certain bumblebee models
generates an NG-sector axial constraint 
with a related remnant gauge freedom
\cite{bk,bfk}.
For simplicity in what follows,
we consider the case of generic $k$.

Once the cardinal gauge \rf{newg} is imposed,
the harmonic condition \rf{harmonic}
follows from the equations of motion.
The latter are found from the Lagrange density \rf{hlag}
to be  
\beq
K_{\mu\nu\al\be} h^{\al\be} 
\equiv -\ka G^L_\mn 
= 0.
\label{heqmot}
\eeq
Contracting these equations in turn with 
$\et^{\mu\nu}$,
$c^{\mu\nu}$,
$(c\et c)^{\mu\nu}$,
and
$(c\et c\et c)^{\mu\nu}$
yields in momentum space the four conditions
\bea
k_\mu h^{\mu\nu}k_\nu 
&=&
0,
\nonumber\\
k_\al c^\al_{\pt{\al}\mu} 
h^{\mu\nu}k_\nu 
&=&
0,
\nonumber\\
k_\al c^\al_{\pt{\al}\be} c^\be_{\pt{\be}\mu} 
h^{\mu\nu}k_\nu 
&=&
0,
\nonumber\\
k_\al c^\al_{\pt{\al}\be} c^\be_{\pt{\be}\ga} 
c^\ga_{\pt{\ga}\mu} 
h^{\mu\nu}k_\nu 
&=&
0.
\label{hilbertreq}
\eea
Collecting the coefficients of 
$h^{\mu\nu}k_\nu$
gives a 4$\times$4 matrix 
that is invertible when Eq.\ \rf{detcond} is satisfied,
which is the case under the present assumptions.
It follows that $h^{\mu\nu}k_\nu= 0$,
and hence that the gauge choice \rf{newg} 
obeys the harmonic condition \rf{harmonic}. 
The equations of motion then reduce to
\beq
\prt^\la \prt_\la h^{\mu\nu} = 0,
\label{redheqmot}
\eeq
and they describe the usual two graviton degrees of freedom
propagating as massless spin-2 waves.

We now have all the ingredients in hand to verify the equivalence
between 
the theory \rf{hlag} for a propagating spin-2 field $h^{\mu\nu}$
and the theory \rf{eplag} for the NG sector of the cardinal model.
Starting with the former,
we can impose the four cardinal gauge conditions \rf{newg}
on the ten independent graviton components $h^{\mu\nu}$.
The equations of motion \rf{heqmot} then imply 
the harmonic condition \rf{harmonic},
which leaves two degrees of freedom that propagate 
as conventional massless modes.
These results are paralleled 
in the theory \rf{eplag} for the NG sector of the cardinal model.
The field $\tg^\mn$ containing the  
Lorentz NG modes $\cE_{\mu\nu}$ 
is subject to the constraints \rf{decid},
so $\tg^\mn$ matches 
the graviton $h^\mn$ in cardinal gauge,
\beq
h^\mn \leftrightarrow \tg^\mn .
\label{lmatch}
\eeq
The harmonic condition holds for both
$\tg^{\mu\nu}$ and $h^{\mu\nu}$.
The equations of motion \rf{deceqmot}
for the Lorentz NG modes $\cE_{\mu\nu}$ 
can be matched directly to 
the equations of motion \rf{heqmot}
for the graviton $h^{\mu\nu}$
by multiplying the latter with $\tO^{\mu\nu\rh\si}$.

Evidently,
the cardinal and graviton theories are in direct correspondence,
even though their gauge structures differ.
The presence of the potential in the linear cardinal theory 
excludes the gauge symmetry of the graviton theory,
but the gauge freedom of the latter
means that only six of the ten components of $h^{\mu\nu}$
are physical or auxiliary,
thereby matching the six Lorentz modes $\cE_{\mu\nu}$ 
in the NG sector of the cardinal theory.
Note also that the gauge freedom of the graviton theory
could be fixed to cardinal gauge in a standard way,
by adding suitable gauge-fixing terms to the Lagrange density.
The parallel in the cardinal theory
would be the presence of Lagrange multipliers
for the constraints \rf{decid}.

\section{Bootstrap procedure}
\label{Bootstrap procedure}

This section considers some generic features
of the bootstrap procedure for
self-consistent coupling to the energy-momentum tensor.
We summarize the Deser version 
\cite{sd}
of the bootstrap for obtaining general relavitity
from the linear graviton theory \rf{hlag},
and we present some generic results
that are useful for the subsequent analysis.

\subsection{Bootstrap for general relativity}
\label{Bootstrap for general relativity}

The analysis takes advantage of the first-order Palatini form 
\cite{ap}
of the nonlinear Einstein-Hilbert action of general relativity,
which can be written as
\bea
S_{\rm GR} &=&
\int d^4 x ~ \ka \fg^\mn R_\mn(\Ga).
\label{gnonlin}
\eea
Here,
$\fg^\mn$ is the tensor density of weight one
defined in terms of the usual reciprocal metric $g^\mn$ as 
\beq
\fg^\mn \equiv \sqrt{|g|} ~g^\mn .
\eeq
Its inverse is a tensor density of weight negative one,
which we define as 
\beq
\fg_\mn \equiv \fr 1 {\sqrt{|g|}} ~g_\mn .
\eeq
Also,
\bea
R_\mn(\Ga) &=&
\prt_\al \gam\al\mu\nu 
- \half \prt_\mu \gam\al\nu\al
- \half \prt_\nu \gam\al\mu\al
\nonumber\\
&&
+ \gam\be\be\al \gam\al\mu\nu 
- \gam\al\mu\be \gam\be\nu\al
\label{curvature}
\eea
is the curvature tensor for the connection
$\gam\al\mu\nu$.
In this approach,
both $\fg^\mn$ and $\gam\al\mu\nu$ 
are viewed as independent fields
at the level of the action,
and the identification of 
$\gam\al\mu\nu$ with the Christoffel symbols
arises on shell 
by solving the equations of motion. 

In what follows,
we define the fluctuation $\fh^\mn$ of $\fg^\mn$ 
about the Minkowski background $\et^\mn$ as
\beq
\fg^\mn = \et^\mn + \fh^\mn.
\eeq
Note the use of contravariant indices 
in this definition.
Also, 
note that $\fh^\mn$ can be identified at linear order
with the usual trace-corrected field $\ol h{}^\mn$:
\beq
\fh^\mn \approx -\ol h{}^\mn \equiv - h^\mn + \half \et^\mn h.
\label{fh}
\eeq

Given the linear graviton theory \rf{hlag},
the nonlinear Einstein-Hilbert action
can be derived 
by adding a coupling
to the energy-momentum tensor $T_{\mu\nu}$
and requiring its conservation be consistent
order by order
\cite{bootstrap}.
Deser has shown that this bootstrap procedure 
can be performed in a single elegant step
\cite{sd}.

The starting point of the derivation is to note that 
the equations of motion \rf{heqmot} for $h^{\mu\nu}$,
obtained in the previous section 
from the second-order Lagrange density \rf{hlag},
also follow from the linearized version 
of the first-order action \rf{gnonlin}.
The latter becomes
\bea
S^L_{\rm GR} &=&
\int d^4 x ~ \cl^L_{\rm GR} ,
\nonumber\\
\cl^L_{\rm GR}
&=&
\ka [\fh^\mn (\prt_\al \gam\al\mu\nu - \prt_\nu \gam\al\mu\al)
\nonumber\\
&&
+ \et^\mn
(\gam\be\be\al \gam\al\mu\nu 
- \gam\al\mu\be \gam\be\nu\al)] ,
\label{hlin}
\eea
with $\fh^\mn$ and $\gam\al\mu\nu$ 
viewed as independent fields.
Variation of $S^L_{\rm GR}$ with respect to these fields
yields two sets of equations of motion.
These fix $\gam\al\mu\nu$ 
as the usual linearized Christoffel symbols,
and they imply the linear equations of motion \rf{heqmot}
for $h^{\mu\nu}$ obtained from the second-order
Lagrange density \rf{hlag}.

The prescription for the bootstrap procedure 
is to require that the energy-momentum tensor $T_{\mu\nu}$
obtained from the action \rf{hlin} is coupled as a source
in a self-consistent manner.
It turns out to be most convenient to work 
with the trace-reversed energy-momentum tensor $\ta_{\mu\nu}$,
which in the linear limit is related to $T_{\mu\nu}$ by
\beq
\ta_{\mu\nu} = T_{\mu\nu} - \half \et_{\mu\nu} T^\al_{\pt{\al}\al}.
\eeq
For a given Lagrange density $\cl$
in Minkowski spacetime with metric $\et_{\mu\nu}$,
the tensor $\ta_{\mu\nu}$
can be calculated via the Rosenfeld method
\cite{lr}.
The procedure involves promoting the Minkowski metric $\et^{\mu\nu}$ 
to an auxiliary weight-one metric density $\ps^{\mu\nu}$ 
and the partial derivative $\prt_\mu$ to the covariant derivative
$D_\mu$ formed using $\ps^{\mu\nu}$,
so that $\cl$ becomes covariant in the auxiliary spacetime.
The trace-reversed energy-momentum tensor $\ta_{\mu\nu}$
is then found from the expression
\bea
-\half \ta_{\mu\nu} &=& 
\fr {\de \cl^L}{\de\ps^{\mu\nu}}\bigg|_{\ps\to\et}.
\label{tadef}
\eea
For the linear theory with Lagrange density $\cl^L_{\rm GR}$,
this yields 
\bea
-\half \ta_{\fh\mn} &=& 
\ka (\gam\be\be\al \gam\al\mu\nu 
- \gam\al\mu\be \gam\be\nu\al) 
+\ka \si_\mn(\fh,\Ga),
\nonumber\\
\eea
where $\si_\mn$ is a total-derivative term given by
\bea
\si^\mn(\fh,\Ga)
&=&
-\half \prt_\ga
\bigl[ \fh^{\mu\ga}\gamu\rh\rh\nu
+\fh^{\nu\ga}\gamu\rh\rh\mu
-\fh^{\mu\nu}\gamu\rh\rh\ga 
\nonumber\\
&&
+\fh^{\mu\rh}(\gamu \nu\rh\ga - \gamu\ga\rh\nu)
+\fh^{\nu\rh}(\gamu \mu\rh\ga - \gamu\ga\rh\mu)
\nonumber\\
&&
\qquad
-\fh^{\ga\rh}(\gamu \mu\rh\nu - \gamu\nu\rh\mu)
\nonumber\\
&&
+\et^{\mn}(\half \tr{\fh\et}\gamu\si\si\ga 
- \fh^{\rh\si}\gaml\rh\si\ga)
\bigr].
\label{simunu}
\eea
On shell,
$\si_\mn$ can be expressed more elegantly as
\beq
\si_\mn = R^L_\mn(\Ga) - R^L_\mn(\Ga^L) ,
\eeq
where $R^L_\mn(\Ga)$ is the linear part of 
the Ricci curvature,
\beq
R^L_\mn(\Ga) =
\prt_\al \gam\al\mu\nu 
- \half \prt_\mu \gam\al\nu\al
- \half \prt_\nu \gam\al\mu\al,
\eeq
and $\Ga^L$ is the linearized Christoffel symbol
\bea
\Ga^L_{\al\mn} &=& 
\half [
\prt_\al\fh_{\mu\nu} 
- \prt_\mu\fh_{\nu\al}
- \prt_\nu\fh_{\mu\al}
\nonumber\\
&&
\quad
+ \half (
\et_{\mu\al}\prt_\nu 
+ \et_{\nu\al}\prt_\mu 
- \et_{\mu\nu}\prt_\al
) \tr{\fh\et} ] .
\quad
\label{lingam}
\eea

The full nonlinear action of general relativity 
is obtained by coupling the nonderivative part of 
$\ta_{\fh\mn}$ as a source for $\fh^\mn$,
\bea
S_{\rm GR} &=&
S^L_{\rm GR} 
+ \int d^4 x ~ \ka \fh^{\mu\nu}
(\gam\be\be\al \gam\al\mu\nu 
- \gam\al\mu\be \gam\be\nu\al).
\nonumber\\
\label{hnonlin}
\eea
Variation of this action with respect to $\fh^\mn$
yields the Einstein equation $R_\mn = 0$ in the form
\beq
\ka R^L_\mn(\Ga) = \half \ta_{\fh\mu\nu} + \si_\mn,
\eeq
which implies
\beq
R^L_\mn(\Ga^L) = 8\pi G_N ~\ta_{\fh\mu\nu}.
\eeq
This verifies that coupling 
the nonderivative part of $\ta_{\fh\mn}$ 
as a source for $\fh^\mn$
indeed produces the usual Einstein equations.
Moreover,
since the nonderivative part of $\ta_{\fh\mn}$ 
is independent of $\et^\mn$,
it generates no additional contribution 
to the energy-momentum tensor
and so no further iteration steps are required.

\subsection{Generic bootstrap results}
\label{Generic bootstrap}

In this subsection,
we outline some generic applications of the bootstrap procedure,
starting from an action given in Minkowski spacetime.
The example relevant in our context 
is either an action $S^{(0)}$ independent of $\fh^\mn$ 
or an action $S^{(1)}$ linear in $\fh^\mn$.
In each case,
we seek to construct the corresponding action $S$
that incorporates consistent self-coupling
to $\fh^\mn$ at all orders.

\subsubsection{Case of $S^{(0)}$}
\label{Case of Szero}

Consider first the case of an action $S^{(0)}$ 
independent of $\fh^\mn$,
such as a matter action.
We write
\beq
S^{(0)} = \int d^4 x ~\cl^{(0)},
\label{Szero}
\eeq
where the Lagrange density
\beq
\cl^{(0)}= \cl^{(0)}(\et_\mn, f_a, \prt_\mu f_a)
\label{Lzero}
\eeq
is a function of the spacetime metric $\et_\mn$,
a set of fields $f_a(x)$, 
and their derivatives $\prt_\mu f_a$.
For the purposes of this work,
it suffices to suppose that 
the terms $\prt_\mu f_a$
are either derivatives of scalars
or are gauge kinetic terms,
so that promotion of $\prt_\mu$ 
to the auxiliary covariant derivative
has no effect:
$\prt_\mu f_a\to D_\mu[\ps] f_a\equiv \prt_\mu f_a$.
This simplifying assumption avoids the need to consider 
terms of the $\si_\mn$ type in the analysis.

To obtain the energy-momentum tensor 
for the action \rf{Szero},
the Lagrange density 
$\cl^{(0)}$
is promoted to a covariant expression
with respect to $\ps^\mn$,
\beq
\cl^{(0)}\to \cl^{(0)}(\ps_\mn, f_a, \prt_\mu f_a).
\label{promote}
\eeq
To ensure $\cl^{(0)}$ remains a density,
multiplication by a factor of a power of $\sqrt{|\ps|}$ 
may be required as part of this promotion,
where $\ps\equiv \det{\ps^\mn}$.
Using the definition \rf{tadef} then yields
\bea
-\half \ta^{(0)}_{\mn} &=& 
\fr {\de \cl^{(0)}}{\de\ps^\mn}\bigg|_{\ps\to\et}.
\label{tauzero}
\eea

The bootstrap procedure requires that $\ta^{(0)}_{\mn}$ 
be consistently coupled as a source for $\fh^\mn$.
The action $S^{(0)}$ must therefore be supplemented 
by an additional term
\bea
S^{(1)} &=& 
\int d^4 x ~ \cl^{(1)}
\equiv
\int d^4 x ~ \fh^\mn (-\half \ta^{(0)}_{\mn}) ,
\label{Sone}
\eea
up to a possible constant.
However,
in general the term $S^{(1)}$ itself contributes 
a term $\ta^{(1)}_{\mn}$ 
to the energy-momentum tensor,
\bea
-\half \ta^{(1)}_{\mn} &=& 
\fr {\de \cl^{(1)}}{\de\ps^{\mu\nu}}\bigg|_{\ps\to\et}
= \fh^\ab 
\fr {\de (-\half \ta^{(0)}_{\ab})}
{\de\ps^\mn}\bigg|_{\ps\to\et}.
\eea
Consistency of the coupling
then requires that a further term $S^{(2)}$ be added
to the action,
\bea
S^{(2)} &=& 
\int d^4 x ~\cl^{(2)},
\eea
where $\cl^{(2)}$ is the solution to the differential equation
\bea
\fr {\de \cl^{(2)}}{\de\fh^{\mu\nu}}\bigg|_{\ps\to\et}
&=& 
-\half \ta^{(1)}_{\mn} 
\equiv 
\fh^\ab \fr {\de (-\half \ta^{(0)}_{\ab})}
{\de\ps^\mn}\bigg|_{\ps\to\et}.
\eea
We find
\bea
\cl^{(2)} 
&=&
\half \fh^\ab \fh^\cd \fr {\de (-\half \ta^{(0)}_{\cd})}
{\de\ps^\ab}\bigg|_{\ps\to\et}
\nonumber\\
&=&
\half \fh^\mn (-\half \ta^{(1)}_{\mn}) ,
\eea
up to a possible constant.

Iterating this procedure yields a series of terms
summing to the desired Lagrange density $\cl$,
\bea
\cl &=& 
\sum_{n=0}^{\infty}
\fr 1 {n!}
\fh^{\al_1\be_1}\cdots \fh^{\al_n\be_n}
\fr {\de^n (-\half \ta^{(0)}_{\al_n\be_n})}
{ \de\ps^{\al_1\be_1}\cdots \de\ps^{\al_{n-1}\be_{n-1}} }
\bigg|_{\ps\to\et}.
\nonumber\\
\eea
The series can be constructed provided the integrability conditions
are satisfied at each step,
and it may terminate at some finite $n$.
It represents a Taylor expansion of $\cl$,
and inspection reveals the identification
\bea
\cl &=&
\cl^{(0)}(\ps_\mn, f_a, \prt_\mu f_a)
\bigg|_{\ps\to\fg}.
\label{identify}
\eea

The above derivation shows that 
knowledge of $\cl^{(0)}$ in the form \rf{Lzero}
suffices to determine $\cl$.
If originally the matter-gravity coupling is specified
in the linearized form \rf{Sone},
the bootstrap procedure amounts to finding $\cl^{(0)}$ 
and then determining $\cl$ via
Eqs.\ \rf{promote} and \rf{identify}. 
If instead a pure matter action is specified
by giving $\cl^{(0)}$,
it suffices to promote it according to Eq.\ \rf{promote}
and obtain $\cl$ via the identification \rf{identify}.
In this case,
the bootstrap corresponds 
to the standard minimal-coupling procedure.
For example,
the usual Minkowski-spacetime energy-momentum tensor
for Maxwell electrodynamics is
\bea
T^{\rm EM}_\mn &=&
F_\mu^{\pt{\mu}\la} F_{\nu\la}
- \frac 1 4 \et_\mn F^\ab F_\ab
= \ta^{(0)}_{{\rm EM}\mn},
\label{emenmomzero}
\eea
with the latter equality following from conformal invariance.
The corresponding Lagrange density is 
\bea
\cl_{\rm EM}^{(0)} &=& 
- \frac 1 4 \et^{\al\ga} \et^{\be\de} F_\ab F_\cd.
\label{emlagzero}
\eea
Promoting this 
according to Eq.\ \rf{promote}
and making the identification \rf{identify}
directly yields the usual Lagrange density $\cl_{\rm EM}$ 
for electrodynamics in curved spacetime,
\bea
\cl_{\rm EM} &=& 
- \fr 1 {4 \sqrt{|\fg|}} 
\fg^{\al\ga} \fg^{\be\de} F_\ab F_\cd,
\label{emlagg}
\eea
where $\fg \equiv \det{\fg^\mn}$.

\subsubsection{Case of $S^{(1)}$}
\label{Case of Sone}

Under some circumstances,
the given starting point is instead an action 
$S^{(1)}$
for a theory linear in $\fh^\mn$.
To obtain the fully coupled action $S$,
one can explicitly perform the iteration procedure above.
However,
a more efficient `inverse' method 
can be adopted instead.
To implement this method,
we start by identifying  
the energy-momentum tensor $\ta^{(0)}_{\mn}$ 
from the specified action $S^{(1)}$
written in the form \rf{Sone},
and we promote it to a covariant expression
with respect to $\ps^\mn$:
\beq
\ta^{(0)}_{\mn} (\et)\to \ta^{(0)}_\mn(\ps).
\label{promotetau}
\eeq
An appropriate multiplicative factor of $\sqrt{|\ps|}$
may be required to maintain the tensor transformation
properties of $\ta^{(0)}_{\mn}$.
We then write the differential equation
\bea
-\half \ta^{(0)}_{\mn} &=& 
\fr {\de \cl^{(0)}}{\de\ps^\mn},
\label{tauzerodiffeq}
\eea
which reduces to Eq.\ \rf{tauzero}
in the limit $\ps^\mn\to\et^\mn$.
The differential equation can be solved
if the integrability condition 
\beq
\fr {\de \ta^{(0)}_{\mn}} {\de\ps^\ab}
= 
\fr {\de \ta^{(0)}_{\ab}} {\de\ps^\mn}
\label{intcondzero}
\eeq
is satisfied.
Once the solution $\cl^{(0)}$ is obtained,
we can apply the identification \rf{identify}
to obtain $\cl$ and hence $S$.

The above inverse trick is applied
in some of the analysis that follows.
To illustrate it in a more familiar context,
consider the cosmological constant $\La$.
In Minkowski spacetime,
$\La$ is associated with
an effective energy-momentum tensor given by
\bea
T^{\La}_\mn &=&
- 2 \ka \La \et_\mn
= - \ta^{(0)}_{\La\mn}.
\label{Laenmom}
\eea
The challenge is to bootstrap this to 
the fully coupled Lagrange density $\cl^\La$.
Following the inverse trick,
we promote $\ta^{(0)}_{\La\mn}$ to
\beq
\ta^{(0)}_{\La\mn} (\ps) = 
2 \ka \La \sqrt{|\ps|} \ps_\mn,
\eeq
where the appropriate factor of $\sqrt{|\ps|}$
has been introduced.
With the identities
\bea
\de\ps_\mn &=& -\ps_{\mu\al}\ps_{\nu\be}\de\ps^\ab,
\nonumber\\
\de\sqrt{|\ps|} &=& \half \sqrt{|\ps|} \ps_\ab \de\ps^\ab,
\eea 
the integrability condition 
\rf{intcondzero} can be verified,
so the differential equation \rf{tauzerodiffeq}
can be solved for $\cl^{(0)}(\ps)$.
Making the identification \rf{identify}
then yields 
\bea
\cl^{\La} &=& 
- 2 \ka \La \sqrt{|\fg|},
\label{cosmconst}
\eea
in agreement with the usual result.
Notice that the linearized version of this is
\bea
\cl^{\La} &\approx& 
- 2 \ka \La - \ka \La \fh^\mn \et_\mn
\nonumber\\
&=& 
- 2 \ka \La + \fh^\mn (-\half \ta^{(0)}_{\La\mn}),
\label{lincosmconst}
\eea
as expected from Eq.\ \rf{Laenmom},
and that the zeroth-order term 
$\cl^{(0)}(\et)$ is merely a constant in this example.
Note also that the first-order term 
$\cl^{(1)}(\et)$ produces a linear instability
in the action at this order.
This could be avoided by initiating the bootstrap 
from a theory formulated in a suitable 
Riemann background spacetime
\cite{sd4}.

As another example,
consider the bootstrap procedure 
for the transverse-traceless (TT) gauge.
A common form for this gauge
involves the trace-corrected field $\ol h{}^\mn$
and a timelike unit vector $n_\mu$:
\beq
\tr{\ol h{}\et} = 0, \quad
n_\mu \ol h{}^\mn = 0, \quad
\prt_\mu \ol h{}^\mn = 0.
\eeq
These standard linear gauge-fixing conditions 
can be expressed in terms of $\fh^\mn$ and 
$\Ga^{L\al}_{\pt{L \al}\mn}$
using Eqs.\ \rf{fh} and \rf{lingam}.
The resulting expressions can then be implemented  
in the linearized action \rf{hlin}
via the addition of the linear Lagrange density
\beq
\cL^L_{\rm TT} =
\la_{(1)} \tr{\fh\et}
+ \la_{(2)\nu} n_\mu \fh^\mn
+ \la_{(3)\al} \et^\mn 
\Ga^{L\al}_{\pt{L \al}\mn} ,
\eeq
where $\la_{(1)}$, $\la_{(2)\nu}$, and $\la_{(3)\al}$
are Lagrange multipliers.
The bootstrap procedure can be applied to each
of the three terms independently.
The first term is linear in $\fh^\mn$ and
of the same form as in Eq.\ \rf{lincosmconst},
so the bootstrap is immediate.
The second term is also linear in $\fh^\mn$,
and the integrability conditions are directly satisified.
The inverse trick described above can therefore be applied.
The third term is independent of $\fh^\mn$,
so the bootstrap method of the previous subsection applies.
The net result of the bootstrap 
is the nonlinear constraint terms
\bea
\cL_{\rm TT} &=&
2 \la_{(1)} (\sqrt{|\fg|} - \sqrt{|\et|}~)
+ \la_{(2)\nu} n_\mu (\fg^\mn - \et^\mn) 
\nonumber\\
&&
+ \la_{(3)\al} \fg^\mn 
\Ga^{\al}_{\pt{\al}\mn} ,
\eea
which correspond to a nonlinear form of the TT gauge constraints,
\bea
\sqrt{|\fg|} = \sqrt{|\et|},
\quad
n_\mu \fg^\mn = n_\mu \et^\mn ,
\quad
\fg^\mn \Ga^{\al}_{\pt{\al}\mn} = 0.
\eea

\section{Bootstrap for cardinal gravity}
\label{Bootstrap for cardinal gravity}

At this stage,
we are in a position to consider 
the nonlinear extension of the cardinal theory
\rf{clag}.
This section begins by presenting 
a convenient first-order reformulation
of the linear cardinal theory.
In this form, 
the bootstrap of the kinetic terms is straightforward
using the methods of the previous section.
We investigate the bootstrap integrability conditions
on an arbitrary potential term,
which turn out to provide interesting constraints
on the theory.
Finally, 
the bootstrap of these terms is also presented.

\subsection{First-order action}

To facilitate comparison with the bootstrap for general relativity,
a first-order form of the theory \rf{clag} is useful.
To develop this,
we introduce the trace-reversed cardinal field $\fC^\mn$ as
\bea
\fC^\mn &=& 
- C^\mn + \half \et^\mn C^\al_{\pt{\al}\al}.
\label{trc}
\eea
Note the signs, 
which are chosen to improve the correspondence
to the conventions used in the analysis of general relativity. 
The field $\fC^\mn$ plays a central role in what follows.

In terms of $\fC^\mn$,
the second-order Lagrange density $\cl_\fC$ yielding equivalent
equations of motion to the theory \rf{clag} takes the form
\bea
\cl_\fC &=&
\half
\fC^{\mu\nu} \fK_{\mu\nu\al\be} \fC^{\al\be} 
- \fV(\fC^{\mu\nu},\et_{\mu\nu}).
\label{fclag}
\eea
Here, 
the quadratic operator $\fK_{\mu\nu\al\be}$ 
is given in cartesian coordinates by 
\bea
\fK_{\mu\nu\al\be} &=&
\frac 1 4 \ka[
- (\et_{\mu\al}\et_{\nu\be} + \et_{\mu\be}\et_{\nu\al}) 
\prt^\la \prt_\la
\nonumber\\
&&
\pt{\frac 1 4 \ka[ }
-\et_{\mu\nu} \prt_\al\prt_\be
-\et_{\al\be} \prt_\mu\prt_\nu
\nonumber\\
&&
\pt{\frac 1 4 \ka[ }
+\et_{\mu\al} \prt_\nu\prt_\be
+\et_{\nu\al} \prt_\mu\prt_\be 
\nonumber\\
&&
\pt{\frac 1 4 \ka[ }
+\et_{\mu\be} \prt_\nu\prt_\al
+\et_{\nu\be} \prt_\mu\prt_\al ].
\label{fclagK}
\eea
Note that acting with this operator 
on the fluctuation $\fh^\ab$
produces the linearized Ricci curvature $R^L_\mn$:
\beq
\fK_{\mu\nu\al\be} \fh^{\al\be} 
\equiv \ka R^L_\mn.
\eeq
Note also that the quantities 
$K_{\mu\nu\al\be} C^{\al\be}$ in Eq.\ \rf{clag} 
and $\fK_{\mu\nu\al\be} \fC^{\al\be}$ 
are related by trace reversal with a sign.
In Eq.\ \rf{fclag},
the potential $\fV(\fC^{\mu\nu},\et_{\mu\nu})$
is determined by the requirement that the equations of motion
\bea
\fK_{\mu\nu\al\be} \fC^{\al\be} - \fr{\de \fV}{\de \fC^{\mu\nu}} 
&=& 0
\label{fceqmot}
\eea
have the same content as the original equations
of motion \rf{ceqmot}.
This requires that
\bea
\fr{\de \fV}{\de \fC^\mn} 
&=& 
-\fr{\de V}{\de C^\mn} 
+\half \et_\mn \et^\ab\fr{\de V}{\de C^\ab} .
\label{fpotcond}
\eea
 
To construct the first-order form of the linear cardinal theory,
we follow a similar path to that 
of the Palatini formalism in general relativity 
discussed in Sec.\ \ref{Bootstrap for general relativity}.
Introducing an independent auxiliary field $\gam\al\mu\nu$,
the Lagrange density \rf{fclag} can be rewritten
in terms of $\fC^\mn$ and $\gam\al\mu\nu$
in the equivalent form
\bea
S^L_\fC &=&
\int d^4 x ~ \cl^L_\fC ,
\nonumber\\
\cl^L_\fC
&=&
\ka [ \fC^\mn (\prt_\al \gam\al\mu\nu - \prt_\nu \gam\al\mu\al)
\nonumber\\
&&
+\et^\mn
(\gam\be\be\al \gam\al\mu\nu 
- \gam\al\mu\be \gam\be\nu\al)] 
+ \fV
\nonumber\\
&\equiv & 
\fK^L + \fV,
\label{fClin}
\eea
where $\fK^L$ is the kinetic part of the Lagrange density. 
Variation of this action with respect to the independent fields
$\fC^\mn$ and $\gam\al\mu\nu$
gives the equations of motion.
With standard manipulations,
the equations of motion 
determine the fields $\gam\al\mu\nu$ 
to be linearized Christoffel symbols of the conventional form 
but depending on $\fC^\mn$ instead of $\fh^\mn$.
They also imply linearized versions
of the equations of motion \rf{fceqmot}
for $\fC^\mn$ obtained from the second-order
Lagrange density \rf{fclag}.

The linearized action \rf{fClin}
can be written in other equivalent forms
by decomposing the cardinal field $\fC^\mn$.
In the minimum of the potential $\fV$,
the field $\fC^\mn$ acquires an expectation value $\fc^\mn$,
\bea
\vev{\fC^\mn} &=& 
\fc^\mn \equiv
- c^\mn + \half \et^\mn c^\al_{\pt{\al}\al}.
\label{lincvev}
\eea
This satisfies the identities 
\bea
\tr{\fc\et} &=& \tr{c\et},
\nonumber\\
\tr{(\fc\et)^2} &=& \tr{(c\et)^2},
\nonumber\\
\tr{(\fc\et)^3} &=& 
- \tr{(c\et)^3}
+\frac 32 \tr{c\et} \tr{(c\et)^2} 
\nonumber\\
&&
-\frac 14 (\tr{c\et})^3 ,
\nonumber\\
\tr{(\fc\et)^4} &=& 
\tr{(c\et)^4}
- 2 \tr{c\et} \tr{(c\et)^3} 
\nonumber\\
&&
+\frac 32 (\tr{c\et})^2 \tr{(c\et)^2} 
-\frac 14 (\tr{c\et})^4 ,
\qquad
\eea
and it also obeys 
\beq
\prt_\al \fc^\mn = 0
\label{fconstc}
\eeq
by virtue of the assumption \rf{constc}.
The fluctuation $\tfC^\mn$ about $\fc^\mn$ is
\bea
\tfC^\mn&=& 
- \tC^\mn + \half \et^\mn \tC^\al_{\pt{\al}\al}.
\label{fflucts}
\eea
The analogue of Eq.\ \rf{flucts} therefore becomes
\bea
\fC^\mn &=& \fc^\mn + \tfC^\mn .
\label{fctildedef}
\eea
An alternative expression for the linearized action \rf{fClin}
is therefore 
\bea
S^L_\tfC &=&
\int d^4 x ~ \cl^L_\tfC,
\nonumber\\
\cl^L_\tfC
&=&
\ka [ \tfC^\mn (\prt_\al \gam\al\mu\nu - \prt_\nu \gam\al\mu\al)
\nonumber\\
&&
+\et^\mn
(\gam\be\be\al \gam\al\mu\nu 
- \gam\al\mu\be \gam\be\nu\al)] 
+ \fV
\nonumber\\
&\equiv & 
\fK_\tfC^L + \fV.
\label{tfClin}
\eea
Note that the two linearized actions 
$S^L_\fC$ and $S^L_\tfC$ are identical,
but by virtue of Eq.\ \rf{fconstc}
the kinetic term $\fK^L$ 
differs from $\fK_\tfC^L$ by a total derivative.
 
The cardinal field $\fC^\mn$ can be further decomposed
into NG modes and massive modes,
in parallel with Eq.\ \rf{decomp}.
We write
\beq
\fC^\mn = \fc^\mn + \ftg^\mn + \ftm^\mn,
\label{linCdecomp}
\eeq
where
the trace-reversed NG field $\ftg^\mn$ is defined as 
\beq
\ftg^\mn = - \tg^\mn + \half \et^\mn \tg^\al_{\pt{\al}\al} 
\label{xtrng}
\eeq
and the trace-reversed massive-mode field is 
\beq
\ftm^\mn = - \tm^\mn + \half \et^\mn \tm^\al_{\pt{\al}\al} .
\label{xtrm}
\eeq
The constraints in the NG sector 
corresponding to Eq.\ \rf{decid} can be written as
\bea
\ttr{\ftg\et (c \et)^j} &=& 0, 
\label{fdecid}
\eea
with $j = 0,1,2,3$,
while the analogue of Eq.\ \rf{arbdemid} is
\beq
\tr{\ftg \et ~F(\fc\et, \ftm\et)} = 0,
\label{fdemid}
\eeq
where $F(\fc\et, \ftm\et)$ is 
an arbitrary matrix polynomial in $\fc\et$ and $\ftm\et$.
Another equivalent form for the action \rf{fClin} 
is therefore
\bea
S^L_{\ftg,\ftm} &=&
\int d^4 x ~ \cl^L_{\ftg,\ftm} ,
\nonumber\\
\cl^L_{\ftg,\ftm}
&=&
\ka [(\ftg^\mn +\ftm^\mn) 
(\prt_\al \gam\al\mu\nu - \prt_\nu \gam\al\mu\al)
\nonumber\\
&&
\quad 
+\et^\mn
(\gam\be\be\al \gam\al\mu\nu 
- \gam\al\mu\be \gam\be\nu\al)] 
+ \fV
\nonumber\\
&&
= \fK^L_{\ftg,\ftm} + \fV,
\label{ftglin}
\eea
where $\fK^L_{\ftg,\ftm}$ denotes the kinetic term
expressed in terms of $\ftg^\mn$, $\ftm^\mn$, 
and $\gam\al\mu\nu$. 

\subsection{Kinetic bootstrap}

With the linear cardinal theory 
massaged into a first-order form 
paralleling that used for general relativity,
we are in a position to investigate the bootstrap
to nonlinear cardinal gravity. 
Since the bootstrap involves adding self-coupling
order by order,
it can be done independently for each part 
in the action.
In particular, 
the bootstrap for the kinetic part 
parallels the bootstrap
for the linearized version \rf{hlin}
of general relativity. 

\subsubsection{Primary bootstrap}

It is perhaps most natural to apply the bootstrap procedure
to the linearized theory in the form \rf{fClin},
which holds prior to the spontaneous Lorentz breaking. 
For the corresponding kinetic term $\fK^L$,
the energy-momentum tensor associated 
with $\fC^\mn$ is of the same form as before,
\bea
-\half (\ta_\fC)_\mn &=& 
\ka (\gam\be\be\al \gam\al\mu\nu 
- \gam\al\mu\be \gam\be\nu\al) 
+\ka \si_\mn,
\qquad
\eea
and the nonlinear kinetic action $S_{\fK, \fC}$
is obtained by coupling its nonderivative part 
as a source for $\fC^\mn$,
\bea
S_{\fK, \fC}
&=&
S_{\fK, \fC}^L
+ \int d^4 x ~ \ka \fC^{\mu\nu}
(\gam\be\be\al \gam\al\mu\nu 
- \gam\al\mu\be \gam\be\nu\al)
\nonumber\\
&=&
\int d^4 x ~ \ka (\et^\mn + \fC^\mn) R_\mn(\Ga) ,
\label{fnonlin}
\eea
where $R_\mn(\Ga)$ is the Ricci curvature defined
via the auxiliary field $\gam\al\mu\nu$ in the usual way, 
\bea
R_\mn(\Ga) &=&
\prt_\al \gam\al\mu\nu 
- \half \prt_\mu \gam\al\nu\al
- \half \prt_\nu \gam\al\mu\al
\nonumber\\
&&
+(\gam\be\be\al \gam\al\mu\nu 
- \gam\al\mu\be \gam\be\nu\al).
\eea
Since the extra term in Eq.\ \rf{fnonlin}
is independent of $\et^\mn$,
no further iteration steps are needed.

In the extremum of the potential $\fV$,
the massive modes vanish and 
the result \rf{fnonlin} for the kinetic bootstrap 
reduces to 
\bea
S_{\fK, \fC}
&\supset&
\int d^4 x ~ \ka (\et^\mn + \fc^\mn + \ftg^\mn) R_\mn(\Ga) .
\label{nglimit}
\eea
The combination $(\et^\mn + \fc^\mn)$ can be viewed 
as playing the role of an effective background metric.
Under a suitable change of coordinates,
this effective metric can be brought to the Minkowski form,
$(\et^\mn + \fc^\mn) \to \et^\mn$.
With the identification 
\beq
\fh^\mn \leftrightarrow \ftg^\mn,
\label{nmatch}
\eeq
which matches the linearized correspondence \rf{lmatch},
it follows that the kinetic action 
$S_{\fK, \fC}$
reduces to the Einstein-Hilbert action
in the limit of vanishing massive modes.
The result \rf{fnonlin} for the kinetic bootstrap 
thereby reveals that the nonlinear cardinal theory
represents an alternative theory of gravity
containing general relativity in a suitable low-energy limit.
The correspondence 
\beq
\fg^\mn \leftrightarrow \et^\mn + \tfC^\mn
\label{match}
\eeq
provides the match between the metric density $\fg^\mn$ 
of general relativity and fields in cardinal gravity.

\subsubsection{Alternative bootstraps}

The derivation of the action 
$S_{\fK, \fC}$ in Eq.\ \rf{fnonlin}
is based on applying the bootstrap
to the linearized cardinal action \rf{fClin}
for the cardinal field $\fC^\mn$.
However,
the spontaneous Lorentz violation
produces a phase transition
that naturally separates the cardinal excitations
into NG and massive modes.
One could therefore instead consider applying the bootstrap
to various choices of excitation 
in the effective theory describing the physics
after the spontaneous symmetry breaking has occurred. 
In the remainder of this subsection,
we consider these alternative bootstrap procedures
and their application to the kinetic term
in the linear cardinal theory. 

Suppose the bootstrap is instead applied
to the alternative linearized cardinal action 
\rf{tfClin}
for the fluctuation $\tfC^\mn$.
This procedure has the possible disadvantage 
of requiring a pre-established value 
for the vacuum expectation $\fc^\mn$.
However,
since $\tfC^\mn$ is a fluctuation,
this procedure does parallel more closely 
the usual bootstrap in general relativity,
for which the relevant field $\fh^\mn$ is also a fluctuation.
The derivation of the nonlinear action $S_{\fK, \tfC}$
from the linearized theory \rf{tfClin}
proceeds as before.
The result for this secondary theory is
\bea
S_{\fK, \tfC}
&=&
S_{\fK, \tfC}^L
+ \int d^4 x ~ \ka \tfC^{\mu\nu}
(\gam\be\be\al \gam\al\mu\nu 
- \gam\al\mu\be \gam\be\nu\al)
\nonumber\\
&=&
\int d^4 x ~ \ka (\et^\mn + \tfC^\mn) R_\mn(\Ga) .
\label{tfnonlin}
\eea
This is equivalent to the action $S_{\fK, \fC}$
under a suitable coordinate transformation. 
We thereby find that the secondary bootstrap 
yields the same physics for the kinetic term
as did the primary bootstrap 
leading to Eq.\ \rf{fnonlin}. 

A tertiary theory could also be countenanced,
in which the bootstrap is applied only to the NG modes $\ftg^\mn$
appearing in the linearized action \rf{ftglin}.
While this procedure also requires a pre-established value 
for the vacuum expectation $\fc^\mn$,
it has the possible advantage of matching
more closely the symmetry structure
of the bootstrap for general relativity.
The key point is that the gauge transformation \rf{gauge},
which fails to be a symmetry of the linearized theory 
due to the potential,
nonetheless does define a symmetry for the pure NG sector
because the potential vanishes for pure NG excitations.
In linearized general relativity,
the analogous gauge symmetry can be related 
to the conserved two-tensor current,
and it morphs into diffeomorphism symmetry
following the bootstrap procedure.
In the present context,
this symmetry structure is reproduced in the pure NG sector
if the bootstrap is applied only to the NG excitation $\ftg^\mn$
in the linearized action \rf{ftglin}.

For this tertiary bootstrap,
the first step is to obtain the energy-momentum tensor
for the kinetic term $\fK^L_{\ftg,\ftm}$
in terms of the NG and massive modes.
The calculations for this step again parallel those 
for the bootstrap in general relativity.
We find
\bea
-\half (\ta_{\ftg,\ftm})_\mn &=& 
\ka (\gam\be\be\al \gam\al\mu\nu 
- \gam\al\mu\be \gam\be\nu\al) 
\nonumber\\
&&
+\ka \si_\mn(\ftg,\Ga) + \ka \si_\mn(\ftm,\Ga),
\eea
where $\si_\mn$ is the total-derivative term
given by Eq.\ \rf{simunu} 
but with modified arguments as indicated.
The prescription for the tertiary bootstrap
is then to couple the nonderivative part of 
$(\ta_{\ftg,\ftm})_\mn$ as a source for $\ftg^\mn$,
\bea
\fK_{\ftg,\ftm} &=&
\fK^L_{\ftg,\ftm}
+ \ka \ftg^{\mu\nu}
(\gam\be\be\al \gam\al\mu\nu 
- \gam\al\mu\be \gam\be\nu\al).
\nonumber\\
\label{ngnonlin}
\eea
This prescription yields the tertiary kinetic action
\bea
S_{\fK,\ftg,\ftm}
&=&
\int d^4 x ~ 
\ka (\et^\mn + \ftg^\mn) R_\mn(\Ga) 
\nonumber\\
&&
\pt{\int d^4 x }
+\ka \ftm^\mn 
(\prt_\al \gam\al\mu\nu - \prt_\nu \gam\al\mu\al).
\label{fkin}
\quad
\eea
Paralleling the case of general relativity,
the extra term in Eq.\ \rf{ngnonlin}
is independent of $\et^\mn$,
so no further iteration steps are needed.
Note that the structure of this result 
implies the auxiliary field $\gam\al\mu\nu$ 
is no longer equivalent on shell to a Christoffel symbol.

The tertiary kinetic action $S_{\fK,\ftg,\ftm}$
differs nontrivially from the primary one $S_{\fK,\fC}$,
and the physical content of the two is also different.
With the identification \rf{nmatch} and in the pure NG sector,
both actions match the Einstein-Hilbert action 
of general relativity.
Their linearized content is also the same 
as that of the linear cardinal theory \rf{clag}.

\subsection{Integrability conditions for potential}

Next,
we investigate the integrability conditions
required to apply the bootstrap 
on the potential term.
We obtain constraints such that $\fV$
obeys the integrability conditions,
and we determine a general form of $\fV$ satisfying
these constraints.

To proceed, 
start with the theory in the form \rf{fclag}
in terms of the cardinal field $\fC^\mn$.
The potential is 
$\fV(\fC^{\mu\nu},\et_{\mu\nu})$,
and it is a scalar.
The only scalars 
that can be formed from $\fC^\mn$ and $\et_\mn$
involve traces of the matrix $\fC\et$.
The scalar $\fX_m$ with $m$ such products has the form
\beq
\fX_m = \tr{(\fC \et)^m}.
\eeq
Since $\fC\et$ is a $4\times 4$ matrix,
only four of these are independent,
so the potential 
$\fV(\fC^{\mu\nu},\et_{\mu\nu})$ 
can be written
\beq
\fV(\fC^{\mu\nu},\et_{\mu\nu}) = 
\fV(\fX_1, \fX_2, \fX_3, \fX_4).  
\eeq
In the minimum of $\fV$,
$\fC^\mn = \fc^\mn$
and the scalars $\fX_m$
have expectation values
\bea
\vev{\fX_m} &=& \tr{(\fc \et)^m} \equiv \fx_m.
\label{fvacsol}
\eea

The next step is to determine the energy-momentum tensor 
$\ta_{\fC\mn}$ 
associated with the potential $\fV$
and check the integrability conditions.
We therefore promote $\fV$
to a covariant expression with respect 
to the auxiliary metric density $\ps^\ab$,
\bea
\fV(\fC^\mn,\et_{\mu\nu}) &\to& 
\sqrt{|\ps|}~\fV(\fC^{\mu\nu}/\sqrt{|\ps|},\sqrt{|\ps|}~\ps_{\mu\nu})
\nonumber\\
&&
= \sqrt{|\ps|}~\fV(\fX_1, \fX_2, \fX_3, \fX_4),
\eea
where the four quantities $\fX_m$ are now
\beq
\fX_m(\ps) = \tr{(\fC \ps)^m}
\eeq
and are scalars with respect to $\ps^\mn$.
In parallel with the bootstrap for the kinetic term,
$\fC^\mn$ is taken to be a tensor density
with respect to $\ps^\mn$
in constructing these expressions.

The energy-momentum tensor $\ta_{\fC\mu\nu}$ is
\bea
-\half \ta_{\fC\mn} &=& 
\fr {\de (\sqrt{|\ps|}~\fV)}{\de\ps^\mn}.
\eea
The bootstrap procedure requires
this to be obtained from an action
by varying with respect to $\fC^\mn$.
We must therefore add to the Lagrange density a term
$\fV^\prime$ 
such that
\bea
\fr {\de \fV^\prime}{\de\fC^\mn}
&=& 
-\half \ta_{\fC\mn} 
= \fr {\de (\sqrt{|\ps|}~\fV)}{\de\ps^\mn}.
\eea
If $\fV^\prime$ is smooth,
then
\bea
\fr {\de^2 \fV^\prime} {\de\fC^\mn\de\fC^\ab}
&=&
\fr {\de^2 \fV^\prime} {\de\fC^\ab\de\fC^\mn},
\eea
which implies
\bea
\fr {\de^2 (\sqrt{|\ps|}~\fV)} {\de\ps^\mn\de\fC^\ab}
&=&
\fr {\de^2 (\sqrt{|\ps|}~\fV)} {\de\ps^\ab\de\fC^\mn}.
\label{integ}
\eea
This is the integrability condition 
for the existence of $\fV^\prime$.
It requires symmetry of the double partial derivative
under the interchange $(\mn)\leftrightarrow (\ab)$.

The double derivative appearing in the result \rf{integ}
can be written as 
\bea
\fr {\de^2 (\sqrt{|\ps|}~\fV)} {\de\ps^\mn\de\fC^\ab}
&=&
\sqrt{|\ps|} 
( A_{m\mn\ab} 
\fV_m 
+ B_{mn\mn\ab} 
\fV_{mn} 
),
\nonumber\\
\eea
where $m$ and $n$ are summed,
with 
\beq
\fV_m 
\equiv \fr {\de \fV} {\de\fX_m},
\quad
\fV_{mn} 
\equiv \fr {\de^2 \fV} {\de\fX_m \de\fX_n},
\eeq
and with the coefficients
$A_{m\mn\ab}$ and $B_{mn\mn\ab}$ 
given by
\bea
A_{m\mn\ab}
&=&
\half \ps_\mn \fr {\de \fX_m} {\de\fC^\ab}
+ \fr {\de^2 \fX_m} {\de\ps^\mn \de\fC^\ab} 
\nonumber\\
&=&
\half m \ps_\mn [ \ps (\fC\ps)^{m-1}]_\ab 
\nonumber\\
&&
- m \sum_{k=0}^{m-1} 
[\ps (\fC\ps)^k]_{\mu\al} 
[\ps (\fC\ps)^{m-1-k}]_{\nu\be} ,
\nonumber\\
B_{mn\mn\ab}
&=&
\half\Big(
\fr {\de \fX_m} {\de\ps^\mn}
\fr {\de \fX_n} {\de\fC^\ab}
+
\fr {\de \fX_n} {\de\ps^\mn}
\fr {\de \fX_m} {\de\fC^\ab}
\Big)
\nonumber\\
&=&
-\half mn 
\Big(
[ \ps (\fC\ps)^m]_\mn [ \ps (\fC\ps)^{n-1}]_\ab 
\nonumber\\
&&
\pt{ \half mn \Big( }
+ [ \ps (\fC\ps)^n]_\mn [ \ps (\fC\ps)^{m-1}]_\ab 
\Big) .
\qquad
\eea
Inspection of these results reveals that the
integrability condition 
is satisfied if and only if 
the combined quantity 
\bea
C_{mn\mn\ab} &=&
\half m \fV_m \ps_\mn [ \ps (\fC\ps)^{m-1}]_\ab 
\nonumber\\
&&
- mn \fV_{mn}
[ \ps (\fC\ps)^m]_\mn [ \ps (\fC\ps)^{n-1}]_\ab 
\qquad
\label{combo}
\eea
is symmetric
under the interchange $(\mn)\leftrightarrow (\ab)$.

Using the Hamilton-Cayley theorem,
we can write
\bea
[ \ps (\fC\ps)^4]_\mn 
&=&
p_1 [ \ps (\fC\ps)^3]_\mn 
- p_2 [ \ps (\fC\ps)^2]_\mn 
\nonumber\\
&&
+ p_3 [ \ps \fC\ps]_\mn 
- p_4 \ps_\mn ,
\eea
where
\bea
p_1 &=& \fX_1,
\nonumber\\
p_2 &=& \half \fX_1^2 - \half \fX_2,
\nonumber\\
p_3 &=& \frac 16 \fX_1^3 - \half \fX_1\fX_2 + \frac 13\fX_3,
\nonumber\\
p_4 &=& \frac 1{24} \fX_1^4 - \frac 14 \fX_1^2 \fX_2 
+ \frac 18 \fX_2^2 + \frac 13 \fX_1 \fX_3 - \frac 14 \fX_4 .
\quad
\eea
Adopting this result
and requiring symmetry of the combination \rf{combo}
reveals that the integrability condition 
imposes the following six partial differential equations
on the potential $\fV$:
\bea
\fV_2 + 8 p_4 \fV_{24} &=& 
- \fV_{11} - 4 p_3 \fV_{14},
\nonumber\\
\frac 32 \fV_3 + 12 p_4 \fV_{34} &=& 
- 2 \fV_{12} + 4 p_2 \fV_{14},
\nonumber\\
2 \fV_4 + 16 p_4 \fV_{44} &=& 
- 3 \fV_{13} - 4 p_1 \fV_{14},
\nonumber\\
- 3 \fV_{13} - 12 p_3 \fV_{34} &=& 
- 4 \fV_{22} + 8 p_2 \fV_{24},
\nonumber\\
- 4 \fV_{14} - 16 p_3 \fV_{44} &=& 
- 6 \fV_{23} - 8 p_1 \fV_{24},
\nonumber\\
- 8 \fV_{24} + 16 p_2 \fV_{44} &=& 
- 9 \fV_{33} - 12 p_1 \fV_{34}.
\label{diffeqs}
\eea

Solutions of these equations that are polynomials in $\fX_m$
can be found by construction,
and they are conveniently classified 
according to the power $q$ of $\fX_1$ appearing in the polynomial.
With some calculation,
we have established that the unique polynomial solutions
for $q\le 4$ are 
\bea
\fY_0 &=& 
1,
\nonumber\\
\fY_1 &=& 
\half \fX_1, 
\nonumber\\
\fY_2 &=& 
\frac 18(\fX_1^2 - 2 \fX_2),
\nonumber\\
\fY_3 &=& 
\frac 1 {48}
( \fX_1^3 - 6 \fX_1\fX_2 + 8 \fX_3 ),
\nonumber\\
\fY_4 &=& 
\frac 1 {384}
(\fX_1^4 - 12 \fX_1^2\fX_2 
+ 12 \fX_2^2 + 32 \fX_1\fX_3 - 48\fX_4 ).
\nonumber\\
\eea
More generally,
it follows that any polynomial
obtained as the term at $O(\fC^q)$ 
in the series expansion
of $\sqrt{|\det{1 + \fC\ps}|}$ is
a solution.
An expression for these polynomials is
\bea
\fY_q &=&
\lim_{\ep\to 0}
\fr 1 {q!}
\fr{\prt^q}{\prt \ep^q} 
(\ep p_1 + \ep^2 p_2 + \ep^3 p_3 + \ep^4 p_4)^{1/2}.
\qquad
\label{polygen}
\eea
For example,
at $q=5$ a solution to Eq.\ \rf{diffeqs} is the polynomial
\bea
\fY_5 &=& 
\frac 1 {768}
(- 3 \fX_1^5 + 28 \fX_1^3\fX_2 - 36 \fX_1 \fX_2^2 
\nonumber\\
&&
\pt{\frac 1 {768}}
- 48 \fX_1^2\fX_3 + 32 \fX_2 \fX_3 + 48 \fX_1 \fX_4 ).
\qquad
\label{genlinpot}
\eea
We conjecture that the polynomials obtained in this way
are in fact unique solutions at each order $q$. 

A general potential $\fV$ that solves
the differential equations \rf{diffeqs} 
can therefore be written as
\bea
\sqrt{|\ps|}~ \fV &=&
\sqrt{|\ps|}~
\sum_{q=0}^{\infty}
\al_q \fY_q,
\label{vcsoln}
\eea
where the $\al_q$ are arbitrary real constants.
For any fixed $\al_q$,
a potential of this form
satisfies the integrability conditions
\rf{integ}
required for the bootstrap procedure.
Note that for the special case $\al_q =\al_0$ for all $q\ge0$,
the solution becomes 
\beq
\sqrt{|\ps|}~ 
\fV = \al_0 \sqrt{|\det{\ps + \fC}|}~.
\eeq

\subsection{Bootstrap for integrable potential}

In this subsection,
we first apply the bootstrap procedure
to the integrable potential \rf{vcsoln}.
We then consider some aspects of extrema
of the resulting theory,
provide a construction for a local minimum,
and offer some remarks
about alternative bootstrap procedures
for the potential.
 
\subsubsection{Potential bootstrap}

The bootstrap procedure 
using the cardinal field $\fC^\mn$
can be explicitly performed term by term
on the potential \rf{vcsoln}.
For each $q$,
$\sqrt{|\ps|}~ \fY_q$ 
is a coefficient in the expansion of  
$\sqrt{|\det{\ps + \fC}|}$.
In Sec.\ \ref{Case of Sone},
a bootstrap procedure has been performed 
that leads to the potential \rf{cosmconst}
proportional to $\sqrt{|\det{\ps + \fC}|}$.
It follows from this analysis that 
the bootstrap applied to the term $\sqrt{|\ps|}~ \fY_q$
generates for each $q$ 
the full result $\sqrt{|\det{\ps + \fC}|}$
minus the sum of all terms of orders less than $q$:
\bea
\sqrt{|\ps|}~ \fY_0
& \to &
\sqrt{|\det{\ps + \fC}|},
\nonumber\\
\sqrt{|\ps|}~ \fY_1
& \to &
\sqrt{|\det{\ps + \fC}|} - \sqrt{|\ps|}~ \fY_0 ,
\nonumber\\
\sqrt{|\ps|}~ \fY_2
& \to &
\sqrt{|\det{\ps + \fC}|} - \sqrt{|\ps|}~ (\fY_0 + \fY_1) ,
\qquad
\eea
and so on,
with the general term being
\bea
\sqrt{|\ps|}~ \fY_q
& \to &
\sqrt{|\det{\ps + \fC}|} - \sqrt{|\ps|}~ \sum_{k=0}^{q-1} \fY_k .
\qquad
\eea

Applying the bootstrap to the general potential \rf{vcsoln}
yields the bootstrap potential $\fV_\fC$,
\bea
\sqrt{|\ps|}~ 
\fV_\fC &=& 
\sum_{q=0}^\infty
\al_q
\Big(
\sqrt{|\det{\ps + \fC}|} - \sqrt{|\ps|}~ \sum_{k=0}^{q-1} \fY_k 
\Big)
\nonumber\\
&=&
\sqrt{|\ps|}~ 
\sum_{q=0}^\infty \al_q 
\sum_{k=q}^{\infty} \fY_k 
\nonumber\\
&=&
\sqrt{|\ps|}~ \sum_{k=0}^\infty \de_k \fY_k ,
\label{fullpotboot}
\eea
where the real coefficients $\de_k$
are given as
\beq
\de_k = \sum_{q=0}^k \al_q.
\eeq
Note that the coefficient $\de_k$ for fixed $k$
acquires nonvanishing contributions
from any nonvanishing coefficients $\al_q$ with $q \le k$.

For nonlinear cardinal gravity,
the above discussion reveals that
the potential term appearing in the bootstrap action
takes the form
\bea
S_{\fV,\fC} &=&
\int d^4x~ \fV_\fC = \sum_{k=0}^\infty \de_k 
\int d^4x~ \fY_k .
\label{fullpotac}
\eea
This potential term combines 
with the kinetic term $S_{\fK,\fC}$
in Eq.\ \rf{fnonlin}
to form the primary cardinal action.

\subsubsection{Extrema of the potential}
\label{Extrema of the potential}
 
Vacuum solutions of nonlinear cardinal gravity 
are extremal solutions of the potential $\fV_\fC$.
In an extremum,
the cardinal field $\fC^\mn$ acquires a vacuum value
that may differ from any extrema  
generated by the potential $\fV$
in the linearized theory
and defined in Eq.\ \rf{lincvev}.
By mild abuse of notation,
in what follows we adopt the same notation $\fC^\mn = \fc^\mn$ 
for a vacuum value in an extremum of $\fV_\fC$.
Similarly,
we adopt the same notation as in Eq.\ \rf{linCdecomp}
for the decomposition of the cardinal field $\fC^\mn$ 
and its fluctuations $\tfC^\mn$
into the NG excitations $\ftg^\mn$ of Eq.\ \rf{xtrng}
and the massive excitations $\ftm^\mn$ of Eq.\ \rf{xtrm}.
However,
linearized results for $\ftg^\mn$ and $\ftm^\mn$
such as Eqs.\ \rf{fdecid} and \rf{fdemid}
no longer hold.

A vacuum of $\fV_\fC$
can also be identified by the values
$\fx_m$ taken by the four scalars $\fX_m$,
as in Eq.\ \rf{fvacsol}.
The restriction of the potential $\fV_\fC$ to the NG sector
can then be achieved by replacing $\fV_\fC$ 
with the Lagrange-multiplier potential 
\beq
\fV_\la = \sum_{m=1}^4 \la_m (\fX_m - \fx_m),
\label{bootlmpot}
\eeq
which excludes fluctuations away from the extremum.
If desired,
the on-shell values of the Lagrange multipliers $\la_m$ 
can be set to zero 
by suitable boundary conditions.
This potential facilitates the identification
of the NG and massive modes.
The NG modes $\ftg^\mn$ are 
the nonzero components of $\fC^\mn$
that preserve the constraints
obtained from the Lagrange-multiplier equations of motion,
while the massive modes are the components of $\fC^\mn$
that are constrained to zero. 
Note that the potential $\fV_\la$ is dynamically equivalent
to a potential $\fV_{\la^\prime}$
expressed using the integrable polynomials \rf{polygen},
given by
\beq
\fV_{\la^\prime}
= \sum_{m=1}^4 \la_m^\prime (\fY_m - \fy_m),
\label{bootlmpotprime}
\eeq
where $\fy_m$ are the values of $\fY_m$
for $\fC^\mn = \fc^\mn$. 
The Lagrange-multiplier constraints are equivalent
by direct comparison,
while the dynamical properties 
under variation with respect to $\fC^\mn$
are equivalent when the Lagrange multipliers
are identified by the nonsingular set of linear equations
\beq
\la_m = \fr{(-1)^{m+1}}{2m}
\sum_{p=m}^4
\la_p^\prime \fy_{p-m}
\eeq
with $1\leq m \leq 4$.

Using the potential \rf{bootlmpot},
the NG modes $\ftg^\mn$
are seen directly to be the solutions of the equations
$\fX_m = \fx_m$,
which can be written as
nonlinear generalizations of Eq.\ \rf{fdecid},
\bea
0 &=& \tr{\ftg \et},
\nonumber\\
0 &=& 2 \tr{\ftg \et c \et}
+ \tr{(\ftg \et)^2},
\nonumber\\
0 &=& 
3 \tr{\ftg \et (c \et)^2} 
+ 3 \tr{(\ftg \et)^2 c \et}
+ \tr{(\ftg \et)^3},
\nonumber\\
0 &=& 
4 \tr{\ftg \et (c \et)^3} 
+ 3 \tr{(\ftg \et)^2 (c \et)^2}
+ 3 \tr{(\ftg \et c \et)^2}
\nonumber\\
&&
+ 4 \tr{(\ftg \et)^3 c \et} 
+ \tr{(\ftg \et)^4}.
\label{nonlinfdecid}
\eea
The ten independent components of $\ftg^\mn$
are constrained by these four equations,
leaving the expected six NG modes.
The four massive modes can be denoted by $\fM_m$
and specified as 
\bea
\fM_m &=& \fX_m - \fx_m
\nonumber\\
&=&
\tr{(\fc \et + \tfC \et)^m} - \tr{(\fc \et)^m} .
\eea
They are contained in the symmetric tensor $\ftm^\mn$,
which is obtained by subtraction of the NG modes $\ftg^\mn$
from the cardinal fluctuation field $\tfC = \fC^\mn-\fc^\mn$.

In the absence of coupling to matter,
the equations of motion for cardinal gravity
are obtained by varying 
the sum of the kinetic and potential actions
\rf{fnonlin} and \rf{fullpotac} 
with respect to the independent fields.
Eliminating the auxiliary field $\gam\al\mu\nu$ 
yields the field equations in the absence of matter as
\bea
R_\mn &=& 2\ka \ta^{\rm vac}_\mn ,
\qquad 
\fX_m = \fx_m,
\eea
where $\ta^{\rm vac}_\mn$ is given by
\bea
-\half \ta^{\rm vac}_\mn &=& 
\fr{\prt\fV_\fC}{\prt\fC^\mn} \Big\vert_{\fC\to \fc}
= \sum_{m=1}^4 \fr{\prt\fX_m}{\prt\fC^\mn}
\fV_{\fC,m}
\Big\vert_{\fC\to \fc}
\nonumber\\
&=& 
\sum_{m=1}^4 m
[\et (\fc\et)^{m-1}]_\mn 
\fV_{\fC,m}\vert_{\fC\to \fc}.
\label{vacenmom}
\eea
Note that $\fV_{\fC,m} = \la_m$ in the Lagrange-multiplier limit.
The quantity $\ta^{\rm vac}_\mn$ 
represents a kind of vacuum energy-momentum tensor density.
Trace-reversing yields 
the field equations for cardinal gravity 
in the absence of matter,
which can be written in the form
\beq
G^\mn = 2\ka T_{\rm vac}^\mn .
\eeq
Here,
$G^\mn$ is the Einstein tensor
for the metric obtained from the metric density 
$(\et^\mn + \fC^\mn)$,
while the vacuum energy-momentum tensor 
$T_{\rm vac}^\mn$
is obtained by the corresponding trace reversal of
$\ta^{\rm vac}_\mn$.
The conservation law
\beq
D_\mu T_{\rm vac}^\mn = 0
\eeq
follows by virtue of the Bianchi identities.
This conservation remains true in the presence of matter couplings,
provided the matter-sector energy-momentum tensor
is independently conserved.
If the Lagrange multipliers $\la_m$ vanish,
or more generally if $\fV_m$ vanishes,
then the vacuum energy-momentum tensor is zero
and the usual form of general relativity is recovered.
Otherwise,
there is a positive or negative contribution 
to the vacuum energy-momentum tensor.
This may play a role in cosmology
and the interpretation of dark energy.
 
In the pure NG sector with zero on-shell Lagrange multiplier fields,
the effective potential vanishes
and nonlinear cardinal gravity 
reduces to the kinetic term \rf{nglimit}.
As already noted,
this limit reproduces general relativity,
with the identification  
$\ftg^\mn\leftrightarrow\fh^\mn$
in Eq.\ \rf{nmatch}.
The Einstein-Hilbert action is recovered in a fixed gauge,
the nonlinear cardinal gauge,
which is defined by the four nonlinear gauge conditions
\bea
0 &=& \tr{\fh \et},
\nonumber\\
0 &=& 2 \tr{\fh \et c \et}
+ \tr{(\fh \et)^2},
\nonumber\\
0 &=& 
3 \tr{\fh \et (c \et)^2} 
+ 3 \tr{(\fh \et)^2 c \et}
+ \tr{(\fh \et)^3},
\nonumber\\
0 &=& 
4 \tr{\fh \et (c \et)^3} 
+ 3 \tr{(\fh \et)^2 (c \et)^2}
+ 3 \tr{(\fh \et c \et)^2}
\nonumber\\
&&
+ 4 \tr{(\fh \et)^3 c \et} 
+ \tr{(\fh \et)^4}
\label{nonlingauge}
\eea
obtained by the replacement $\ftg^\mn\to\fh^\mn$
in Eq.\ \rf{nonlinfdecid}.

The bootstrap for general relativity
transforms the gauge symmetry \rf{hgauge}
of the linearized theory 
into diffeomorphism invariance of the Einstein-Hilbert action,
involving particle transformations 
of the metric density $\fg^\mn$.
In the linear cardinal theory,
the analogue of the gauge symmetry \rf{hgauge}
is the symmetry \rf{gauge} of the kinetic term alone.
The pre-bootstrap potential $\fV$ explicitly breaks this symmetry,
so the potential term \rf{fullpotac}
can be expected to exhibit diffeomorphism breaking 
under particle transformations of 
the analogue metric density $(\et^\mn+\fC^\mn)$.
This is reflected,
for example,
in the presence of a factor $\sqrt{|\ps|}\to \sqrt{|\et|}=1$
in the measure of Eq.\ \rf{fullpotac}.
However,
as expected from the match to general relativity,
the pure NG sector of cardinal gravity 
with zero on-shell Lagrange multipliers 
does exhibit the usual diffeomorphism invariance 
because the potential vanishes in this sector.
Note also that 
cardinal gravity remains invariant
under diffeomorphisms of the Minkowski spacetime. 
 
Both general relativity and 
cardinal gravity are invariant under
(observer) general coordinate transformations.
The match between the two theories in the pure NG limit
involves a coordinate transformation 
taking $(\et^\mn + \fc^\mn) \to \et^\mn$
in the kinetic term \rf{nglimit}.
There is therefore a corresponding transformation
taking $\et^\mn \to [(1 + \fc\et)^{-1} \et]^\mn $
in the potential term. 
For example,
the general coordinate invariance ensures a factor
$\sqrt{|(1+\fc\et)^{-1}\et|}$ 
appears in the measure of Eq.\ \rf{fullpotac}.
However,
the vanishing of the potential in the pure NG sector
makes this factor irrelevant for the match to general relativity.

\subsubsection{Stability of the extrema}

Given a bootstrap potential $\fV_\fC$,
an interesting issue is whether it admits
an extremum that is stable.
The question of overall stability 
for any given theory with Lorentz violation is involved 
\cite{kleh}.
Even for the comparatively simple bumblebee theories
the issue remains open,
although considerable recent progress has been made
\cite{stability}.
A full analysis for cardinal gravity 
lies outside the scope of this work.
Instead,
this subsection provides a few remarks
on stability in the specific context of the potential term.

In the vacuum,
the extremal solutions obey
\bea
0 &=& \fr {\prt\fV_{\fC}}{\prt \fC^\mn}
\Big\vert_{\fC\to \fc}
\nonumber\\
&=& 
\sum_{m=1}^4
m [ \et (\fc\et)^{m-1}]_\mn \fV_{\fC,m} 
\Big\vert_{\fC\to \fc},
\eea
where 
$\fV_{\fC,m} \equiv \prt\fV_\fC/\prt \fX_m$.
By assumption,
the matrix $\fc\et$ has four inequivalent nonzero eigenvalues.
Working in the basis in which $\fc\et$ is diagonal,
this implies 
the generic conditions for a vacuum are
\beq
\fV_{\fC,m}
\Big\vert_{\fC\to \fc}
= 0.
\eeq

A vacuum of $\fV_{\fC}$ is stable
if it is a Morse critical point 
with positive definite hessian.
For simplicity,
we introduce the explicit diagonal basis
\beq
\fC^{\mu\la}\et_{\la\nu} = \fC_\mu \de^\mu_{\pt{\mu}\nu}
\eeq
(no sum on $\mu$),
where the four quantities $\fC_\mu$ are the 
eigenvalues of $\fC\et$.
Then
\beq
\fX_m = \sum_{j = 0}^3 (\fC_j)^m,
\eeq
and in the vacuum $\fC_j = \fc_j$,
with all four values $\fc_j$ inequivalent and nonzero.
In the diagonal basis,
stability depends on the hessian
\bea
H_{jk} &=&
\fr {\prt^2 \fV_\fC}
{\prt \fC_j \prt \fC_k}
\Big\vert_{\fC\to \fc}
\nonumber\\
&=&
\sum_{m,n=1}^{4}
m n (\fc_j)^{m-1} (\fc_k)^{n-1}
\fV_{\fC,mn}
\Big\vert_{\fC\to \fc} .
\nonumber\\
\eea
If the discriminant is nonzero and the four eigenvalues
$H_m$
of the hessian are positive,
the extremum is a local minimum.

An analytical derivation of a potential
with a positive definite hessian
in terms of the polynomial basis \rf{polygen}
is challenging.
Instead,
we proceed by ansatz 
using the shifted variables
\beq
\tfX_m = \fX_m - \fx_m.
\eeq
For the ansatz,
we adopt the form of a Taylor expansion
\bea
\fV_\fC = 
\half a_{mn} \tfX_m \tfX_n
+ \frac 1 6 a_{mnp} \tfX_m \tfX_n \tfX_p
+\ldots,
\label{ansatz}
\eea
where the coefficients $a_{mn}$, $a_{mnp}$, $\ldots$
are real constants.
The potential $\fV_\fC$ in Eq.\ \rf{fullpotac}
is a combination of integrable partial potentials,
so the expression \rf{ansatz} must be integrable too.
We can therefore constrain the coefficients
by imposing the integrability conditions \rf{diffeqs}
on $\fV_\fC$ itself at $\tfX_m = 0$.
At second order in $\tfX_m$,
this imposes six conditions 
on the ten degrees of freedom $a_{mn}$.
The four degrees of freedom $a_{m4}$
can be taken as unconstrained at this order.
To impose the integrability conditions at third order,
it is convenient to take partial derivatives
of Eqs.\ \rf{diffeqs}
with respect to each $\fX_m$.
This produces 24 equations,
which combine with the second-order equations
to yield 16 independent constraints
on the 20 third-order coefficients $a_{mnp}$.
The four degrees of freedom $a_{m44}$ 
can be taken as unconstrained at this order.
Proceeding in this way,
we find a $4(n-1)$-dimensional solution space
for the potential $\fV_\fC$ up to order $n$.
As a check,
the resulting solutions can be reconstructed
in terms of suitable combinations 
of the polynomial basis \rf{polygen}.

Given the potential $\fV_\fC$ in the form \rf{ansatz},
the issue of finding 
a solution with positive definite hessian
can be resolved numerically.
Investigation shows that there is a subspace
of coefficients $a_{mn}$
for which the integrability conditions are satisfied
and the hessian is positive definite.
An explicit example is the potential
\beq
\fV_\fC = \sum_{k=1}^8 \de_k \fY_k ,
\eeq
with the coefficients given by 
\bea
\de_1 &\simeq& -2.81,
\quad
\de_2 \simeq -5.46,
\quad
\de_3 \simeq 13.1,
\quad
\de_4 \simeq 19.3,
\nonumber\\
\de_5 &\simeq& -24.7,
\quad
\de_6 \simeq -29.6,
\quad
\de_7 \simeq 16.0,
\quad
\de_8 \simeq 17.1.
\nonumber\\
\eea 
The local minimum is found to lie at
\beq
\fX_1 \simeq 0.250,
\quad
\fX_2 \simeq 2.06,
\quad
\fX_3 \simeq 0.578,
\quad
\fX_4 \simeq 1.44.
\eeq
The eigenvalues of the corresponding hessian 
are found to be
\beq
H_1 \simeq 2.80,
\quad
H_2 \simeq 0.927,
\quad
H_3 \simeq 0.104,
\quad
H_4 \simeq 0.0579 ,
\eeq
demonstrating positivity.
This example therefore represents 
a potential $\fV_\fC$ having a local minimum.

\subsubsection{Alternative potential bootstraps}

The bootstrap procedure discussed above holds
for the potential 
prior to the development of a vacuum value
for the cardinal field $\fC^\mn$.
Alternative options for the potential term,
applicable following spontaneous Lorentz violation instead,
include a secondary bootstrap
using the cardinal fluctuation $\tfC^\mn$
and a tertiary one using only the NG modes $\ftg^\mn$.
The explicit construction of these potentials
lies outside the scope of this work.
Instead,
this subsection contains a few brief comments 
about some aspects of these alternative bootstrap procedures,
following from the analysis of the primary case.

To perform an alternative bootstrap procedure,
the corresponding integrable potential 
must first be constructed.
For the secondary bootstrap 
involving the cardinal fluctuation $\tfC^\mn$
introduced in Eq.\ \rf{fctildedef},
the promotion of the potential $\fV$
to a covariant expression 
with respect to the auxiliary metric density $\ps^{\al\be}$
involves the four scalars $\fX_m$ given by
\bea
\fX_m (\ps) &=& \tr{(\fc \ps + \tfC \ps)^m}.
\eea
The energy-momentum tensor must now be obtained
from an action by varying with respect to $\tfC^\mn$.
The basic integrability condition is found to be 
\bea
\fr {\de^2 (\sqrt{|\ps|}~\fV)} {\de\ps^\mn\de\tfC^\ab}
&=&
\fr {\de^2 (\sqrt{|\ps|}~\fV)} {\de\ps^\ab\de\tfC^\mn}.
\label{tinteg}
\eea
However,
since the cardinal fluctuation $\tfC^\mn$ is merely 
a constant shift of the cardinal field $\fC^\mn$,
we have
\beq
\fr{\prt \fX_m} {\prt \tfC^\mn} =
\fr{\prt \fX_m} {\prt \fC^\mn}.
\eeq
This in turn means that 
the integrability condition is satisfied
for the same symmetry requirement on the same expression \rf{combo}
as before.
The integrable potential for the secondary bootstrap
therefore takes the same form \rf{vcsoln}
as for the primary case.

A similar situation holds for the tertiary bootstrap
involving the NG modes $\ftg^\mn$
in the decomposition \rf{linCdecomp}.
In this case,
the four relevant scalars are 
\bea
\fX_m &=& 
\tr{(\fc \ps + \ftg \ps + \ftm \ps)^m}.
\eea
The energy-momentum tensor is required to arise
by varying an action with respect to $\ftg^\mn$.
This generates the integrability condition
\bea
\fr {\de^2 (\sqrt{|\ps|}~\fV)} {\de\ps^\mn\de\ftg^\ab}
&=&
\fr {\de^2 (\sqrt{|\ps|}~\fV)} {\de\ps^\ab\de\ftg^\mn}.
\label{nginteg}
\eea
However,
the form of Eq.\ \rf{linCdecomp} implies 
\beq
\fr{\prt \fX_m} {\prt \ftg^\mn} =
\fr{\prt \fX_m} {\prt \fC^\mn}.
\eeq
It follows that the integrability condition is again satisfied
for the same symmetry requirement on the same expression \rf{combo},
and the integrable potential for the tertiary bootstrap
takes the same form \rf{vcsoln} as before.

Although the integrable potentials \rf{vcsoln} are the same, 
the alternative bootstrap procedures differ
from each other and from the primary one presented above.
Moreover,
performing these bootstrap procedures 
involves additional choices 
because integration with respect 
to the linear cardinal fluctuation or the linear NG modes
can either be continued at all orders
or can be adjusted at each order to incorporate
the induced nonlinearities.
Any of these bootstrap procedures could in principle be performed
using the methods presented in Sec.\ \ref{Bootstrap procedure}.

An extremum of an alternative bootstrap potential
is achieved for vanishing massive modes.
It can therefore be represented by 
a suitable Lagrange-multiplier potential.
In particular,
in the pure NG limit the potential vanishes 
for on-shell multipliers,
and so the resulting effective theory 
is controlled by the corresponding kinetic term.
This means that general relativity 
is also recovered in the low-energy limits 
of the nonlinear theories arising
in these alternative bootstrap procedures.

\section{Coupling to matter}
\label{Coupling to matter}

At the linear level,
the cardinal field $C^\mn$ must couple to other fields 
in the Minkowski spacetime via a symmetric two-tensor current.
Given our gravitational interpretation of the cardinal field,
the other fields in the theory can be regarded as the matter.
They provide one natural two-tensor current,
the energy-momentum tensor $T_{{\rm M}\mn}$
in the Minkowski spacetime.
We can therefore expect the linearized theory \rf{clag}
to incorporate the matter interaction 
\bea
\cl^L_{{\rm M},C} &=& 
\half C^\mn T_{{\rm M}\mn}.
\label{Cmattint}
\eea
No coupling constant is necessary for this interaction,
since it can be absorbed in the scaling factor $\ka$
already present in the original theory \rf{clag}.

\subsection{Primary bootstrap}

The bootstrap procedure involving the cardinal field $\fC^\mn$
can be applied to the matter interaction \rf{Cmattint}
to determine the form of the matter coupling
for cardinal gravity.
For this purpose,
the interaction \rf{Cmattint}
is conveniently expressed in terms
of the trace-reversed energy-momentum tensor
$\ta_{{\rm M}\mn}$ for the matter.
This tensor arises by variation of the Lagrange density
$\cl_{\rm M}$ for the matter fields via
\bea
-\half \ta_{{\rm M}\mn} &=&
\fr {\de \cl_{\rm M}(\et\to\ps)}{\de\ps^\mn}\bigg|_{\ps\to\et}
\eea
in the usual way.
We can therefore write
\bea
\cl^L_{{\rm M},\fC} &=& 
-\half \fC^\mn 
\ta_{{\rm M}\mn} 
\label{mcint}
\eea
for the matter interaction with the cardinal field $\fC^\mn$.

To perform the bootstrap, 
the techniques of Sec.\ \ref{Generic bootstrap}
can be applied.
The Lagrange density \rf{mcint} is linear in $\fC^\mn$
and so has the form \rf{Sone},
for which the bootstrap yields Eq.\ \rf{identify}.
The bootstrap therefore generates the Lagrange density
\bea
\cl_{{\rm M},\fC} &=& 
\sqrt{|\et+\fC|} ~\cl^L_{{\rm M},\fC}\big|_{\et\to\et+\fC}.
\label{cbootmlag}
\eea

Some insight into the physical content 
of this result 
can be obtained by expanding
about an extremum of the bootstrap potential.
Writing $\fC^\mn = \fc^\mn$ in the extremum 
and denoting the corresponding fluctuations
by $\tfC^\mn = \ftg^\mn + \ftm^\mn$ 
as before,
we obtain
\bea
\cl_{{\rm M},\fC} &=& 
\sqrt{|\et+\fc+\tfC|} ~\cl^L_{{\rm M},\fC}\big|_{\et\to\et+\fc+\tfC}.
\label{ctcbootmlag}
\eea
A comparison of this result 
to the matter coupling of general relativity 
can be performed
by adopting Lagrange-multiplier bootstrap potential \rf{bootlmpot}.
The massive modes vanish,
$\ftm^\mn \to 0$,
and as before a suitable change of coordinates must be performed
to implement the transformation $(\et^\mn + \fc^\mn) \to \et^\mn$
and thereby ensure the kinetic term \rf{nglimit}
contains the conventional Minkowski metric.
The resulting Lagrange density
$\cl^{\rm NG}_{{\rm M},\fC}$
then matches the usual matter term 
$\cl^{\rm GR}_{\rm M}$
in general relativity,
\bea
\cl^{\rm NG}_{{\rm M},\fC} &=&
\sqrt{|\et+\ftg|} ~\cl^L_{{\rm M},\fC}\big|_{\et\to\et+\ftg}
\nonumber\\
&\leftrightarrow& 
\cl^{\rm GR}_{\rm M}
= \sqrt{|\fg|} ~\cl^L_{\rm M}\big|_{\et\to\fg},
\eea
when the correspondence
$\fg^\mn \leftrightarrow \et^\mn + \ftg^\mn$ of Eq.\ \rf{match}
is adopted.

We can therefore conclude that the pure NG sector of cardinal gravity 
with zero on-shell Lagrange multipliers
exactly reproduces general relativity,
including the matter coupling.
When the massive modes are included,
the matter coupling deviates from that in general relativity
by terms that are suppressed by the scale 
of the massive modes.

\subsection{Alternative bootstraps}

Alternative bootstrap procedures 
for the matter coupling can be countenanced instead.
We consider here the secondary and tertiary procedures
discussed above for the kinetic and potential terms.
We also examine some experimental implications of the results
for the pure NG sector and the match to general relativity.

The secondary bootstrap involving 
the cardinal fluctuation $\tfC^\mn$
starts from the matter coupling \rf{Cmattint}
in the form
\bea
\cl^L_{{\rm M},\tfC} &=& 
\fc^\mn (-\half\ta_{{\rm M}\mn}) 
+ \tfC^\mn (-\half\ta_{{\rm M}\mn}) .
\qquad
\label{mtcint}
\eea
The bootstrap can be performed using the methods
of Sec.\ \ref{Generic bootstrap}.
The first term in Eq.\ \rf{mtcint}
involves $\fc^\mn$ but is independent of $\tfC^\mn$,
while the only dependence on the Minkowski metric
appears in $\ta_{{\rm M}\mn}$.  
The effect of the bootstrap on this term is therefore
to introduce a factor of $\sqrt{|\et + \tfC|}$
and to replace 
$\ta_{{\rm M}\mn}(\et^\mn)$ with
$\ta_{{\rm M}\mn}(\et^\mn + \tfC^\mn)$.
The second term is linear in $\tfC^\mn$
and hence is of the form \rf{Sone},
for which the bootstrap gives Eq.\ \rf{identify}.
We therefore obtain 
\bea
\cl_{{\rm M},\tfC} &=& 
\sqrt{|\et+\tfC|} ~\fc^\mn 
(-\half\ta_{{\rm M}\mn} \big|_{\et\to\et+\tfC})
\nonumber\\
&&
+\sqrt{|\et+\tfC|} ~\cl^L_{{\rm M},\fC}\big|_{\et\to\et+\tfC}
\label{tcbootmlag}
\eea
as the secondary bootstrap matter coupling.

For the tertiary bootstrap,
the starting point is the matter coupling in the form
\bea
\cl^L_{{\rm M},\ftg,\ftm} &=& 
(\fc^\mn + \ftm^\mn) (-\half\ta_{{\rm M}\mn}) 
+ \ftg^\mn (-\half\ta_{{\rm M}\mn}). 
\nonumber\\
\label{mngint}
\eea
Here,
we bootstrap only the field $\ftg^\mn$ 
containing the linearized NG modes,
without correcting at each order.
Using the techniques in Sec.\ \ref{Generic bootstrap},
we find the Lagrange density
\bea
\cl_{{\rm M},\ftg,\ftm} &=& 
\sqrt{|\et+\ftg|} ~(\fc^\mn + \ftm^\mn) 
(-\half\ta_{{\rm M}\mn} \big|_{\et\to\et+\ftg})
\nonumber\\
&&
+\sqrt{|\et+\ftg|} ~\cl^L_{{\rm M},\ftg,\ftm}\big|_{\et\to\et+\ftg}
\label{ngbootmlag}
\eea
as the result of the tertiary booststrap.

The alternative results \rf{tcbootmlag} and \rf{ngbootmlag}
for the matter coupling
contain terms corresponding to the usual minimally coupled 
Lagrange density for matter
and additional couplings 
between between matter and the massive modes.
Each also contains a term 
involving the cardinal vacuum value $\fc^\mn$
and the energy-momentum tensor.
This last term remains as an unconventional expression 
in the Lagrange density 
in the pure NG limit $\ftm^\mn \to 0$,
and for the match to general relativity
it therefore represents an unconventional contribution
to the matter sector.

Couplings involving tensor vacuum values appear naturally in the 
Standard-Model Extension (SME),
which provides a general framework 
for the description of Lorentz violation
using effective field theory
\cite{akgrav,ckp}.
The matter sector of the SME includes
Lorentz-violating operators controlled by coefficients
that are symmetric observer two-tensors 
and that can be related to $c^\mn$. 
Numerous experimental measurements have been performed
on the coefficients for Lorentz violation
\cite{tables}.
This offers an interesting opportunity
to identify constrains on the alternative bootstrap theories.

Consider first an example
illustrating the connection between the cardinal matter coupling
and the SME framework,
involving a matter Lagrange density 
for a complex scalar field $\ph$
in Minkowski spacetime given by
\bea
\cl^0_\ph &=&
- \et^\mn \prt_\mu \ph^\dagger \prt_\nu\ph
- U(\ph^\dagger\ph) .
\eea
Here,
$U(\ph^\dagger\ph)$ 
is an effective Lorentz-invariant potential
that can include mass and self-interaction terms.
The corresponding energy-momentum tensor $T^0_\mn$ is
\bea
T^0_\mn =
\prt_{\mu}\ph^\dagger \prt_{\nu}\ph
+ \prt_{\nu}\ph^\dagger \prt_{\mu}\ph
+ \et_\mn \cl_\ph^0 .
\eea
Introducing the cardinal coupling \rf{Cmattint}
and restricting attention to the vacuum value $c^\mn$ 
adds the term
\beq
\cl^\ph_c = \half c^\mn T^0_\mn = \fc^\mn (-\half \ta^0_\mn),
\eeq
where $\ta^0_\mn$ is the trace-reversed form of $T^0_\mn$. 
Performing either of the alternative bootstraps in the NG limit
yields the contribution 
of the cardinal-scalar coupling to the full theory,
\bea
\cl^\ph_{\fc,\ftg} &=& 
\sqrt{|\fg|} ~[\fc^\mn (-\half \ta^0_\mn)]
\big|_{\et\to\fg}
\nonumber\\
&=&
\half \sqrt{|g|} ~ c^\mn T^0_\mn \big|_{\et\to g}
\nonumber\\
&=&
\half \sqrt{|g|} ~ c^{{\rm T}\mn} T^0_\mn \big|_{\et\to g}
\nonumber\\
&&
+ \frac 1 8 \sqrt{|g|} ~ \tr{c g} \tr{T^0 \big|_{\et\to g}g},
\label{phiboot}
\eea
where we denote the bootstrap metric density 
$\et^\mn+\ftg^\mn$ by $\fg^\mn$
and the corresponding metric by $g_\mn$. 
For the last expression in this equation, 
the coefficient $c^\mn$ has been separated into
traceless and trace pieces
for convenience in what follows,
via the definitions
\beq
c^\mn = c^{{\rm T}\mn} + \frac 1 4 \tr{c g} g^\mn,
\quad
\tr{c^{\rm T} g} = 0.
\eeq 

We can compare the result for $\cl^\ph_{\fc,\ftg}$
to that obtained in the
SME framework for the Lorentz-violating theory
of a complex scalar field in Riemann spacetime
with Lagrange density \cite{akgrav}
\bea
\cl^\ph_g &=& 
- \sqrt{|g|} ~g^\mn \prt_\mu \ph^\dagger \prt_\nu\ph
- \sqrt{|g|} ~U(\ph^\dagger\ph) 
\nonumber\\
&&
+ \half \sqrt{|g|} ~k^\mn 
(\prt_{\mu}\ph^\dagger \prt_{\nu}\ph +
\prt_{\nu}\ph^\dagger \prt_{\mu}\ph) .
\label{phimodel}
\eea
In this model,
$k^\mn$ is a symmetric coefficient for Lorentz violation,
which is normally taken to satisfy $\tr{k g} = 0$
because a nonzero trace is Lorentz invariant.
Inspection reveals the identification 
\beq
c^{{\rm T}\mn} \equiv k^\mn 
\label{idck}
\eeq
between the cardinal vacuum value 
and the SME coefficient for Lorentz violation.
Note that the conformally invariant case satisfies 
$\tr{T^0 g} = 0$,
in which case the two models
\rf{phiboot} and \rf{phimodel} match exactly.

As another example with direct physical application,
consider 
the Maxwell Lagrange density 
$\cl_{\rm EM}^{(0)}$ 
for photons in Minkowski spacetime,
given in Eq.\ \rf{emlagzero}.
The corresponding energy-momentum tensor $T^{\rm EM}_\mn$ 
is presented in Eq.\ \rf{emenmomzero}.
The cardinal-photon coupling is
\beq
\cl^\ph_c = \half c^\mn T^{\rm EM}_\mn 
= \fc^\mn (-\half \ta^{\rm EM}_\mn),
\eeq
and the bootstrap generates the result
\bea
\cl^{\rm EM}_{\fc,\ftg} &=& 
\sqrt{|\fg|} ~[\fc^\mn (-\half \ta^{\rm EM}_\mn)]
\big|_{\et\to\fg}
\nonumber\\
&=&
\half \sqrt{|g|} ~ c^{{\rm T}\mn} 
F_\mu^{\pt{\mu}\al} F_{\nu\al}.
\label{EMboot}
\eea
In this example,
only the trace part $c^{{\rm T}\mn}$ appears in the final answer
because the photon action is conformally invariant.
This result can be compared to 
the CPT-even part of the photon sector
in the minimal SME
\cite{km}.
The corresponding coefficients for Lorentz violation
form an observer four-tensor $(k_F)^{\al\la\mu\nu}$,
which has the symmetries of the Riemann tensor.
This four-tensor can be decomposed in parallel with 
the decomposition of the Riemann tensor
into the Weyl tensor, the tracless Ricci tensor, 
and the scalar curvature.
The scalar part is Lorentz invariant.
The Weyl part involves an observer four-tensor
that controls birefringence of light induced by Lorentz violation.
The traceless Ricci part determines the anisotropies
in the propagation of light due to Lorentz violation,
and it is specified by the traceless observer two-tensor
$k_F^\mn \equiv (k_F)^{\al}_{\pt{\al}\mu\al\nu}$.
Only the latter effects are relevant for present purposes.
Restricting attention to these coefficients
produces in Riemann spacetime the Lagrange density
\cite{akgrav}
\bea
\cl_{\rm EM} 
&=& -\frac 1 4 \sqrt{|g|} ~F^\mn F_\mn
+\frac 1 2 \sqrt{|g|} ~k_F^\mn F_\mu^{\pt{\mu}\al} F_{\al\nu}
\nonumber\\
&=& -\frac 1 4 \sqrt{|g|} ~F^\mn F_\mn
+\frac 1 2 \sqrt{|g|} ~k_F^\mn T^{\rm EM}_\mn ,
\eea
where the tracelessness of $k_F^\mn$ has been used.
Comparison of this result with Eq.\ \rf{EMboot}
shows the match
\beq
c^{{\rm T}\mn} \equiv k_F^\mn,
\label{idck2}
\eeq
in analogy with that of Eq.\ \rf{idck}.

The similarity of the matches \rf{idck} and \rf{idck2}
between $c^{{\rm T}\mn}$ and certain traceless
SME coefficients for Lorentz violation
is no accident.
Consider a theory in which the spacetime metric 
in the gravity sector is $g_\mn$.
If the theory has Lorentz violation,
the matter-sector metric could differ from $g_\mn$.
Denote the matter-sector metric by $g_\mn + k_\mn$,
where the coefficient $k_\mn$ for Lorentz violation 
is symmetric and traceless.
For small $k_\mn$,
the matter-sector Lagrange density $\cl_{\rm M}(g+k)$
can be expanded as
\bea
\cl_{\rm M} (g+k) &=&
\cl_{\rm M} (g) 
+ k_\mn \fr {\de \cl_{\rm M}(g)}{\de g^\mn}
+ \ldots
\nonumber\\
&=&
\cl_{\rm M} (g) 
+ \half k_\mn T_{\rm M}^\mn + \ldots ,
\eea
where 
$T_{\rm M}^\mn$ is the energy-momentum tensor for  
the Lagrange density $\cl_{\rm M}(g)$.
We see that the piece of the cardinal coupling \rf{Cmattint}
involving $c^{{\rm T}\mn}$ 
can always be matched at leading order 
to a term involving a traceless shift $k_\mn$
in the matter-sector metric
of a theory with Lorentz violation.

The same line of reasoning also yields a path
to experimental constraints on $c^{{\rm T}\mn}$.
The key point is that
a suitable choice of coordinates
can convert $g_\mn + k_\mn \to g^\prime_\mn$,
thereby making the matter sector Lorentz invariant
at leading order in $k_\mn$.
The price for this transformation 
is the conversion of the gravity-sector metric 
$g_\mn \to g^\prime_\mn - k_\mn$,
which means that signals from Lorentz violation
could be detectable in suitable gravitational experiments.
In particular,
at leading order we find
\bea
\cl_{\rm cardinal} &\supset& 
\ka \fg^\mn R_\mn(\Ga)
\nonumber\\
&\to &
\ka \fg^{\prime\mn} R_\mn(\Ga) + \ka k^\mn R_\mn(\Ga).
\eea
The last term matches the standard form 
for one type of Lorentz violation in the gravity sector
of the minimal SME,
controlled by the coefficient
$s^\mn$ for Lorentz violation
\cite{akgrav}.
This coefficient can be studied experimentally
in various ways
\cite{qbak,akjt}.
Most components of related coefficients have been constrained to 
parts in $10^5$ to $10^{10}$
via reanalysis of several decades of data from lunar laser ranging
\cite{bcs}
and by laboratory tests with atom interferometry
\cite{muller}.
We can therefore conclude that 
the traceless part of the vacuum value of the cardinal field
is constrained at the same level
in both the secondary and the tertiary cardinal theories.

\section{Summary and Discussion}
\label{Discussion}

This work constructs an alternative theory of gravity,
which we call cardinal gravity,
based on the idea that gravitons are massless NG modes 
originating in spontaneous Lorentz violation. 
The starting point is the simple theory \rf{clag}
of a symmetric two-tensor cardinal field $C^\mn$
in Minkowski spacetime
with a potential triggering spontaneous Lorentz violation
\cite{kp}.
Requiring consistent self-coupling
to the energy-momentum tensor
constrains the form of the potential
to the form \rf{vcsoln}.
It also defines a bootstrap procedure
that permits the construction 
of a self-consistent nonlinear theory.

When the bootstrap is applied to the original theory
prior to the spontaneous Lorentz violation,
cardinal gravity emerges.
This theory has kinetic term $S_{\fK, \fC}$ given by
Eq.\ \rf{fnonlin},
potential term $S_{\fV,\fC}$ given by
Eq.\ \rf{fullpotac},
and matter coupling $\cl_{{\rm M},\fC}$ given by
Eq.\ \rf{cbootmlag}.
At low energies compared to the scale of the massive modes,
the potential can be approximated by its extremal
Lagrange-multiplier form \rf{bootlmpot} 
that allows only NG excitations about the vacuum.  
In this limit,
the nonlinear cardinal action reduces to 
the Einstein-Hilbert action of general relativity
with conventional matter coupling
and possibly a vacuum energy-momentum term \rf{vacenmom},
all expressed in the nonlinear cardinal gauge 
given by Eq.\ \rf{nonlingauge}.

If instead the bootstrap is applied 
to the effective action for the spontaneously broken theory,
alternative cardinal theories are generated.
Using the fluctuation field about the cardinal vacuum value
as the basis for the bootstrap yields
a secondary cardinal gravity.
This has kinetic term given by
Eq.\ \rf{tfnonlin}
and matter coupling given by
Eq.\ \rf{tcbootmlag}.
Using instead only the NG excitations to perform the bootstrap 
produces a tertiary cardinal gravity,
with kinetic term given by
Eq.\ \rf{fkin}
and matter coupling given by
Eq.\ \rf{ngbootmlag}.
The actions of these alternative cardinal theories
also reduce to the Einstein-Hilbert action
in the pure NG limit
and in the nonlinear cardinal gauge \rf{nonlingauge}.
However,
unconventional matter coupling terms remain
in this limit.
These can be constrained by suitable gravitational experiments,
and existing results limit
the magnitude of components of the cardinal vacuum value 
to parts in $10^5$ to $10^{10}$. 

All forms of cardinal gravity 
differ from general relativity in certain respects. 
One is the presence of the massive modes $\ftm^\mn$.
The scale of these modes is set by the curvature
of the potential about the Lorentz-violating extremum.
The natural scale in the theory is the Planck mass,
which enters via the Newton gravitational constant 
in the usual way,
so it is plausible that the fluctuations 
of the modes $\ftm^\mn$ are also of Planck mass.
At low energies,
their propagation can therefore be neglected,
and they can be integrated out of the action 
to yield their effective contribution.
The form of the kinetic term \rf{fnonlin}
suggests the corrections to the Einstein-Hilbert action
appear in part as the square of the Ricci tensor
suppressed by the square of the mass of the modes $\ftm^\mn$.
A suppressed effective matter self-interaction
that is quadratic in the energy-momentum tensor
also appears.
Investigation of the resulting subleading corrections
to the Einstein equations,
some of which are proportional to the Ricci tensor
and hence vanish in the vacuum,
is an open topic.
A post-newtonian study of the experimental consequences 
for laboratory and solar-system situations,
including gravitational-wave searches,
would be of definite interest.
A study of the implications for cosmology
would also be worthwhile
because corrections appear to standard solutions
and also because the vacuum energy-momentum tensor \rf{vacenmom}
can appear.
These various investigations may be most effectively undertaken
in the nonlinear cardinal gauge \rf{nonlingauge},
for which the form of conventional general-relativistic solutions
remains to be obtained. 

In more extreme situations,
such as near the singularities of black holes 
or in the very early Universe,
the contributions from the massive modes
could be sufficient to change qualitatively 
the usual general-relativistic behavior.
The additional propagating modes can be expected
to affect features such as inflation
and to change the cosmic gravitational background.
At sufficiently high temperatures
the potential changes shape 
\cite{dj}
to restore exact Lorentz symmetry,
with an extremum having a zero value for $C_{\mu\nu}$.
This reverse phase transition
converts the NG modes into massive modes,
so the graviton excitations acquire Planck masses
and the nature of gravity 
at the big bang is radically changed.

Cardinal gravity has general coordinate invariance
and diffeomorphism symmetry of the background spacetime 
at all scales,
as discussed in the context of 
the gauge-fixing conditions \rf{nonlingauge}.
Diffeomorphism invariance 
involving the analogue metric density $(\et^\mn + \ftg^\mn)$
emerges in the low-energy limit,
where the match to general relativity occurs.
This feature of cardinal gravity has some appeal.
The aesthetic and mathematical advantages
of the diffeomorphism invariance of general relativity
are maintained in the low-energy limit of cardinal gravity,
while at high energies 
the presence of the original background spacetime 
may offer conceptual and calculational advantages 
for understanding the physics.
One example might involve the vacuum value of the metric,
which is presumably set by processes at the Planck scale.
In general relativity
one can ask why the vacuum value of the metric is nonzero.
Since the metric is the fundamental field
and the Einstein-Hilbert action has diffeomorphism invariance,
it might seem natural for the metric field to vanish in the vacuum.
In contrast,
in cardinal gravity at high energies
the background spacetime is nondynamical,
and the gravitational properties at high energies 
are controlled instead by the cardinal field.
The vacuum value of the cardinal field
affects the physics
but not the existence of spacetime properties.
Another example might be improved prospects 
for quantum calculations at high energies,
although this would require revisiting
the analysis in the present work with quantum physics in mind.
For instance,
our derivation of the integrable potential
is based on purely classical considerations,
and the effect of radiative corrections is an open issue.
In the context of bumblebee theories,
requiring one-loop stability under the renormalization group
restricts the form of the potential 
and shows that those producing spontaneous Lorentz breaking
are generic
\cite{bak}.
The analogue of this for cardinal gravity 
represents an independent condition on the potential
that is likely to constrain further its form.

We conclude this discussion by noting 
an interesting possibility implied by the present work. 
We have demonstrated here that nonlinear gravitons
in general relativity 
can be interpreted as NG modes
from spontaneous Lorentz violation.
It is also known that photons can be interpreted
as NG modes from spontaneous Lorentz violation,
even in the presence of gravity:
the Einstein-Maxwell equations are reproduced 
at low energies by a suitable bumblebee theory
\cite{bk}.
Both the graviton and the photon have
two physical propagating modes.
However,
spontaneous Lorentz violation
and the accompanying diffeomorphism violation
can generate up to ten NG modes 
\cite{bk},
so the possibility exists in principle
of developing a combined cardinal-bumblebee theory 
in which the graviton and the photon simultaneously emerge 
as NG modes from spontaneous Lorentz violation.
This would represent an alternative unified framework 
for understanding the long-range forces in nature. 

\section*{Acknowledgments}

This work was supported in part
by the United States Department of Energy
under grant DE-FG02-91ER40661
and by the
Funda\c{c}\~ao para a Ci\^encia e a Tecnologia
in Portugal.


\begin{thebibliography}{99}
\bibitem{foot1}
The terminology refers to the cardinal numbers.
General relativity and cardinal gravity
match at low energies but differ at the Planck scale.
This is in loose analogy 
to the ordinal and cardinal numbers,
which correspond for finite sets
but differ for transfinite ones.
\end{thebibliography}
\end{document}